\RequirePackage{fix-cm}

\documentclass[smallcondensed
]{svjour3}     
\smartqed  
\usepackage{graphicx}
\usepackage{epstopdf}
\usepackage{geometry}
\usepackage{comment}
\usepackage{xspace}
\usepackage{amsmath}
\usepackage{amssymb}

\usepackage{epsfig}
\usepackage{subfigure}
\usepackage{algorithmic}
\usepackage{graphics}
\usepackage{alltt}
\usepackage{float}
\usepackage{array}
\usepackage{footmisc}
\usepackage{comment}
\usepackage{algorithm}
\usepackage{url}
\usepackage[misc]{ifsym}

\usepackage[justification=centering]{caption}

\begin{document}

\title{Efficient Region of Visual Interests Search for Geo-multimedia Data
}


\author{Chengyuan Zhang  $^\dagger$\and
        Yunwu Lin $^\dagger$\and
        Lei Zhu $^\dagger$\and
        Zuping Zhang $^\dagger$\and
        Yan Tang $^\dagger$\and
        Fang Huang $^\dagger$\and
        }


\institute{Chengyuan Zhang \at
            \email{cyzhang@csu.edu.cn}
            \and
            Yunwu Lin \at
              \email{lywcsu@csu.edu.cn}
           \and
           \Letter Lei Zhu \at
              \email{leizhu@csu.edu.cn}
           \and
           Zuping Zhang \at
              \email{zpzhang@csu.edu.cn}
           \and
           Yan Tang \at
              \email{tangyan@csu.edu.cn}
           \and
           Fang Huang \at
              \email{hfang@csu.edu.cn}
           \and
           $^\dagger$ School of Information Science and Engineering, Central South University, PR China\\
}

\date{Received: date / Accepted: date}

\maketitle

\begin{abstract}
With the proliferation of online social networking services and mobile smart devices equipped with mobile communications module and position sensor module, massive amount of multimedia data has been collected, stored and shared. This trend has put forward higher request on massive multimedia data retrieval. In this paper, we investigate a novel spatial query named region of visual interests query (RoVIQ), which aims to search users containing geographical information and visual words. Three baseline methods are presented to introduce how to exploit existing techniques to address this problem. Then we propose the definition of this query and related notions at the first time. To improve the performance of query, we propose a novel spatial indexing structure called quadtree based inverted visual index which is a combination of quadtree, inverted index and visual words. Based on it, we design a efficient search algorithm named region of visual interests search to support RoVIQ. Experimental evaluations on real geo-image datasets demonstrate that our solution outperforms state-of-the-art method.


\end{abstract}

\section{Introduction}
\label{intro}

In the past decade, we have witnessed the rapid development of Internet techniques such as online social networking services, search engine and multimedia sharing services, which generate massive amount of multimedia data~\cite{DBLP:conf/cikm/WangLZ13,DBLP:journals/tip/WangLWZZH15}, e.g., text, image, audio and video. For example, for online social networking services, the most famous social networking site, Facebook (https://facebook.com/), reports 350 million images uploaded everyday in the end of November 2013. On the popular social networking services, Twitter (http://www.twitter.com/), more than 400 million tweets with texts and images have been generated by 140 million users. In September 2017, the largest online social networking platform in China, Weibo (https://weibo.com/) have 376 million active users and more than 100 million micro-blogs are posted by them. For multimedia data~\cite{DBLP:journals/corr/abs-1804-11013} sharing services, More than 3.5 million new images uploaded to Flickr(https://www.flickr.com/) which is the most famous photos sharing web site everyday in March 2013. every minute there are 100 hours of videos which are uploaded to YouTube (https://www.youtube.com/), and more than 2 billion videos totally stored in this platform by the end of 2013. In China, the total watch time monthly of the largest online video service, IQIYI (http://www.iqiyi.com/), exceeded 42 billion minutes and the number of users is more than 230 million monthly. The total amount of audio had exceeded 15 million in Himalaya (https://www.ximalaya.com/), the very popular audio sharing platform in China as of December 2015. These web services not only not only provide great convenience for our daily life, but creates possibilities for the generation, storage and sharing of large-scale multimedia data~\cite{DBLP:conf/mm/WangLWZZ14,DBLP:journals/cviu/WuWGHL18}. Moreover, this trend has put forward higher request on massive multimedia data retrieval~\cite{DBLP:conf/mm/WangLWZ15,DBLP:journals/tip/WangLWZ17,DBLP:conf/sigir/WangLWZZ15,DBLP:conf/ijcai/WangZWLFP16}.

\begin{figure*}
\newskip\subfigtoppskip \subfigtopskip = -0.1cm
\begin{minipage}[b]{0.99\linewidth}
\begin{center}
     \includegraphics[width=1\linewidth]{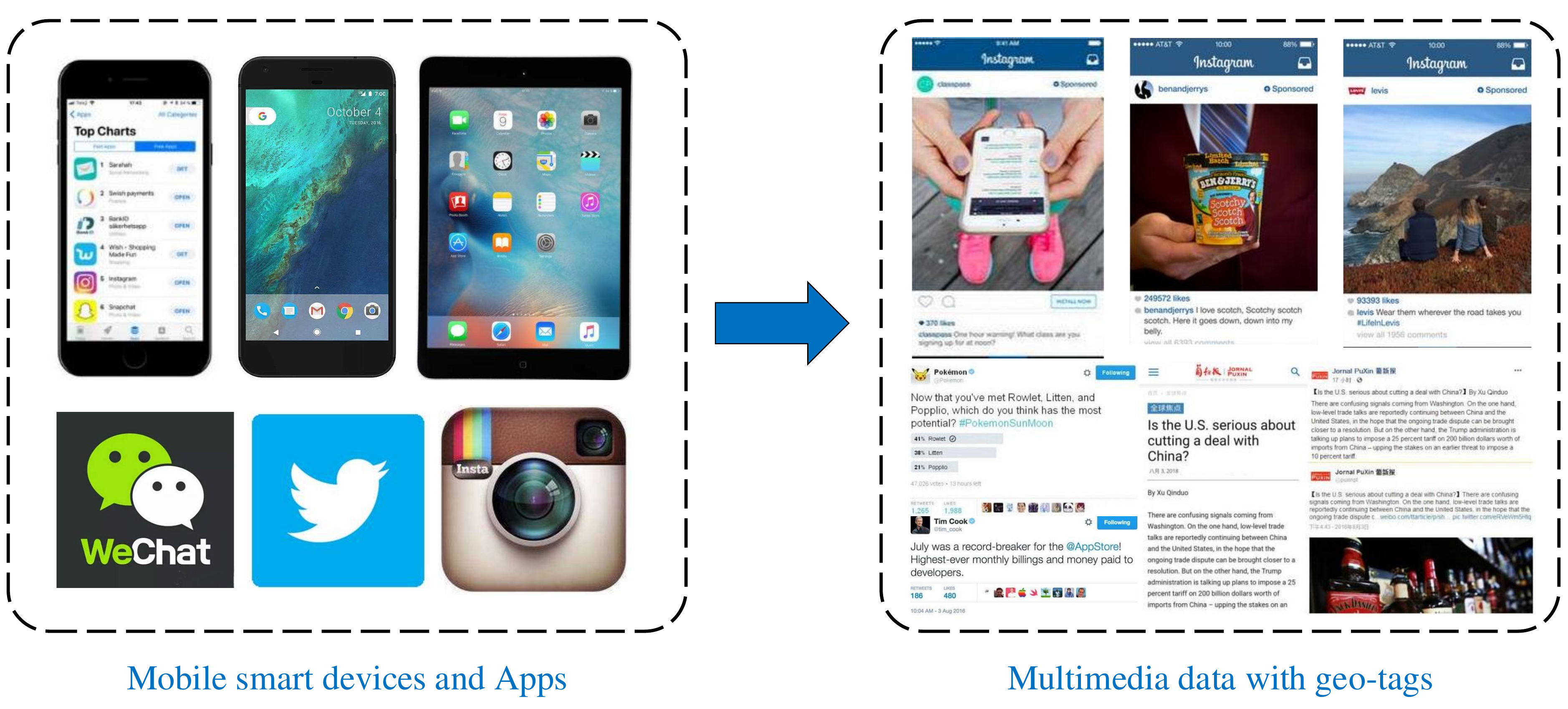}
   \captionsetup{justification=centering}
       \vspace{-0.2cm}
\caption{Mobile smart devices collective multimedia data with geo-tags}
\label{fig:fig1}
\end{center}
\end{minipage}
\label{fig:k}
\end{figure*}

As shown in Figure.~\ref{fig:fig1}, mobile smart devices equipped with mobile communications module (e.g., WiFi and 4G module) and position sensor module (e.g., GPS-Module) such as smartphones and tablets collective huge amounts of multimedia data~\cite{DBLP:journals/corr/abs-1708-02288,DBLP:conf/pakdd/WangLZW14,DBLP:journals/ivc/WuW17} with geo-tags. For example, users can take photos or videos~\cite{LINYANGARXIV} with the geo-location information of the shoot place. Many mobile applications such as WeiChat, Twitter and Instagram support uploading and sharing of images and text with geo-tags. Other location-based services such as Google Places, Yahoo!Local, and Dianping provide the query services for geo-multimedia data by taking into account both geographical proximity and multimedia data similarity. Users can find the place they want to go by these query services, such as \emph{where is the nearest restaurant serving steak and seafood}, \emph{which shop has this type of hat and scarf within a kilometre range}. This type of query can be named geo-multimedia query.

\begin{figure*}
\newskip\subfigtoppskip \subfigtopskip = -0.1cm
\begin{minipage}[b]{0.99\linewidth}
\begin{center}
     \includegraphics[width=1\linewidth]{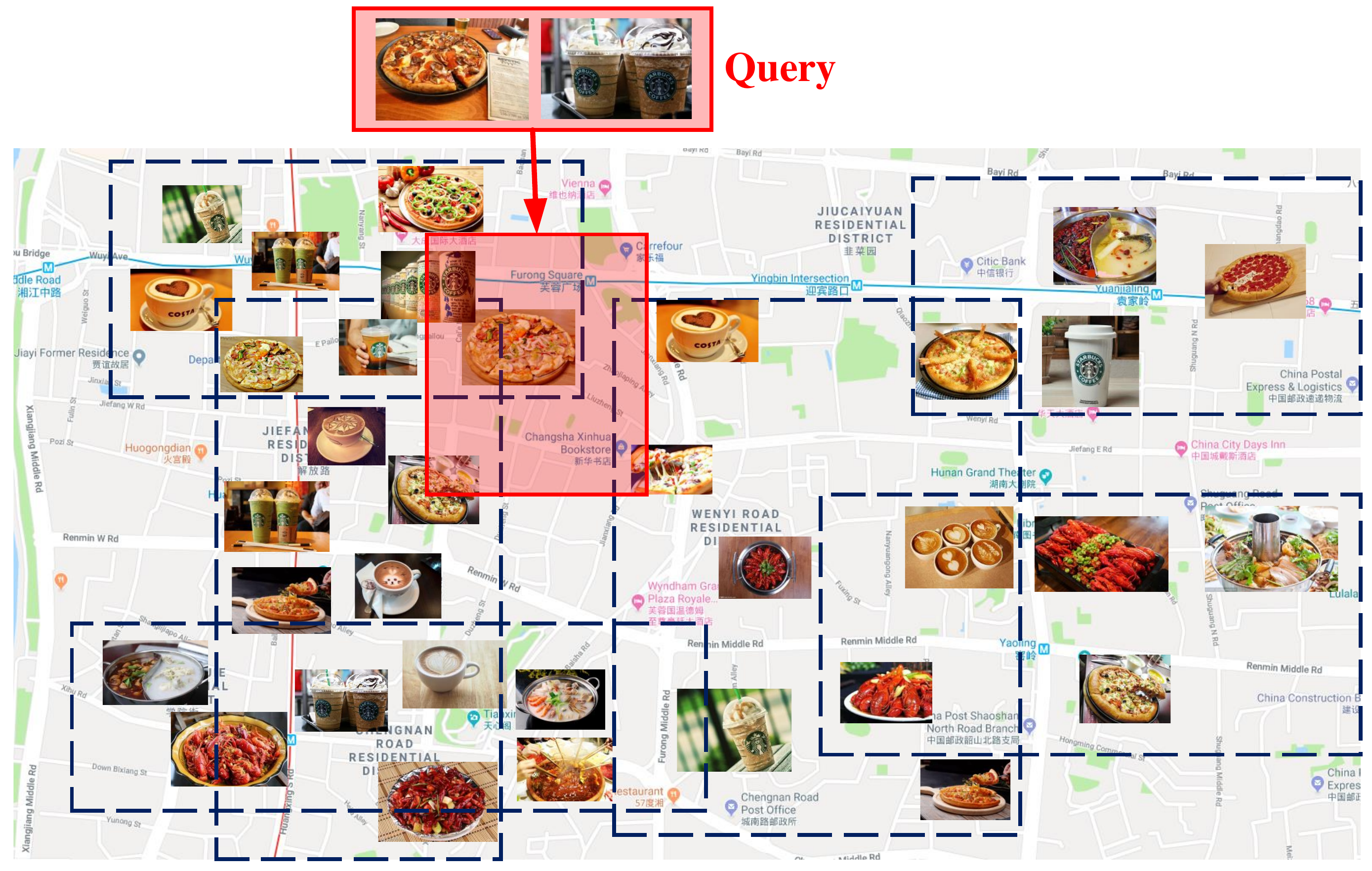}
   \captionsetup{justification=centering}
       \vspace{-0.2cm}
\caption{An example of region of visual interest query}
\label{fig:fig2}
\end{center}
\end{minipage}
\label{fig:k}
\end{figure*}

\noindent\textbf{Motivation}. Spatial keyword search problem has become a hot issue in the community of spatial database due to the mobile applications and location-based services. this type of problem is to find out spatial objects considering two aspects of relevance, i.e., geo-location proximity and textual similarity. Many spatial indexing techniques have been developed like R-Tree~\cite{DBLP:conf/sigmod/Guttman84}, R$^*$-Tree~\cite{DBLP:conf/sigmod/BeckmannKSS90}, IL-quadtree~\cite{DBLP:journals/tkde/ZhangZZL16}, KR$^*$-Tree~\cite{DBLP:conf/ssdbm/HariharanHLM07}, IR$^2$-Tree~\cite{DBLP:conf/icde/FelipeHR08} etc. Deng et al.~\cite{DBLP:journals/tkde/DengLLZ15} studied a generic
version of closest keywords search called best keyword Cover. Cao et al.~\cite{DBLP:conf/sigmod/CaoCJO11} proposed the problem of collective spatial keyword querying, Fan et al.~\cite{Fan2012Seal} studied the problem of spatio-textual similarity search for regions of interests query. However, these researches just only consider the textual data such as keywords and spatial information, they do not take into account other multimedia data mentioned above, like images. Therefore, these solutions cannot be applied in the problem of geo-multimedia data query. In this paper, we aim to design a efficient solution to overcome the challenge of geo-multimedia query. We in the first time propose a novel geo-multimedia query called region of visual interest query. Figure.~\ref{fig:fig2} is an simple but intuitive example to describe this query problem.

\begin{example}
\label{ex:example1}
As illustrated in Figure.~\ref{fig:fig2}, an user who works for a chain restaurant which mainly serves pizza and coffee. She want to investigate which block the people who like pizza and coffee reside in. This can guide advertising in a specific area or district in the way of Internet advertising push service. As more and more people like sharing his or her favorite food by photos with the geo-location information of the shoot place in the Internet, the user can submit a region of visual interests which contains serval images about pizza and coffee and a geographical information of a region. Then the system will return a set of users which are high relevant to the query in both aspects of geographical proximity and visual content similarity.
\end{example}

To our best knowledge, we are the first to propose the problem of region of visual interests (RoVI) query. To solve this problem effectively and efficiently, we present the definition of region of visual interests query and the relevant notions. We introduce how to exploit existing techniques to address the RoVI query problem by present three baselines namely double index, visual first index, and spatial first index. After that, a novel spatial indexing structure is proposed named quadtree based inverted visual index which is a combination of quadtree, inverted index and visual words. Based on this indexing technique, we design a novel efficient search algorithm named region of visual interests search to improve the performance of search.

\noindent\textbf{Contributions}. Our main contributions can be summarized as follows:
\begin{itemize}
\item To the best of our knowledge, we are the first to propose region of visual interests query. Firstly we introduce the definition of geo-image and region of visual interests query and relevant notions. The visual similarity function and geographical similarity function are proposed.
\item We introduce three hybrid indexes namely double index, visual first index, and spatial first index to explain how to exploit existing techniques to address the problem mentioned above.
\item To solve the region of visual interests query problem efficiently, we present a novel spatial indexing structure named quadtree based inverted visual index, which combines quadtree, inverted index and visual words. Based on it, we develop an efficient search algorithm called region of visual interests search (RoVI Search) for this problem.
\item We have conducted extensive experiments on real geo-image dataset. Experimental results demonstrate that our solution outperforms the state-of-the-art method.
\end{itemize}

\noindent\textbf{Roadmap.} In the remainder of this paper, In Section~\ref{relwork} we present the related works about spatial keyword query and image retrieval. In Section~\ref{preliminaries} we propose the definition of region of visual interests query and related conceptions. We introduce three baselines namely double index, visual first index, and spatial first index in Section~\ref{base}. In Section~\ref{spatialindex} we propose a novel spatial indexing technique named quadtree based inverted visual index and an efficient search algorithm named region of visual interests search. In Section~\ref{perform} we present the experiment results. Finally, we conclude the paper in Section~\ref{con}.

\section{Related Work}
\label{relwork}

In this section, we introduce the related works of spatial keywords query and content-based image retrieval, which are related to our work.

\noindent\textbf{Spatial Keywords Query.} Recently, spatial keywords query has become a hot spot attracting many researchers in the filed of spatial database. A spatial keyword query aims to return a set of spatial textual objects which are relevant to the query in both aspects of spatial proximity and textual similarity~\cite{DBLP:conf/er/CaoCCJQSWY12}. Lots of efficient
spatial indexing techniques have been presented such as R-Tree~\cite{DBLP:conf/sigmod/Guttman84}, R$^*$-Tree~\cite{DBLP:conf/sigmod/BeckmannKSS90}, IR-Tree~\cite{DBLP:journals/pvldb/CongJW09}, IR$^2$-Tree. Jo˜ao B. Rocha-Junior~\cite{DBLP:conf/ssd/RochaGJN11} et al. presented a novel spatial indexing structure called Spatial Inverted Index (S2I) to solve top-$k$ spatial keyword query problem. This index maps each distinct term to a set of objects containing the term. They designed two efficient algorithms named \emph{SKA} and \emph{MKA} based on S2I to enhance the performance of top-$k$ spatial keyword search. Wang et al.~\cite{DBLP:conf/icde/WangZZLW15} studied the problem of processing a large amount of continuous spatial keyword queries over streaming data. They proposed a novel adaptive index called AP-Tree which can adaptively groups registered queries using keyword and spatial partitions. Li et al.~\cite{DBLP:journals/tkde/LiLZLLW11} developed a new spatial indexing structure named IR-Tree, which is together with a top-$k$ document search algorithm facilitates four tasks for document searches problem. Zhang et al.~\cite{DBLP:conf/edbt/ZhangTT13} proposed a scalable integrated inverted index called $I^3$ which used quadtree to hierarchically partition the data space into cells. Zheng et al.~\cite{DBLP:conf/icde/ZhengSZSXLZ15} investigated interactive top-$k$ spatial keyword (IT$k$SK) query and they designed a three-phase solution focusing on both effectiveness and efficiency. To solve the problem of top-$k$ spatial keyword search, a novel index structure called IL-Quadtree was proposed by Zhang et al.~\cite{DBLP:journals/tkde/ZhangZZL16}. This technique is developed to use both spatial and keyword to effectively reduce the search space.

Other spatial keyword search problems have been proposed and well studied. Deng et al.~\cite{DBLP:journals/tkde/DengLLZ15} proposed a novel spatial keyword search problem named Best Keyword Cover, and they presented a novel scalable algorithm named keyword nearest neighbor expansion. Guo et al.~\cite{DBLP:conf/sigmod/GuoCC15} proved that the problem of $m$CK
queries is NP-hard. In addition, they propose a 2-approximation greedy approach as a baseline and designed two approximation algorithms called \emph{SKECa} and \emph{SKECa+} to solve this problem efficiently. João B. Rocha-Junior et al.~\cite{DBLP:conf/edbt/Rocha-JuniorN12} solved the top-$k$ spatial keyword queries problem on road networks. Dislike the spatial search problem mentioned above, the geographical distance between the query location and an object is the shortest path in road network. They developed new spatial indexing structures and algorithms that are able to solve this peoblem efficiently. Lee et al.~\cite{DBLP:journals/tkde/LeeLZT12} developed a novel system framework called \emph{ROAD} to solve the problem of spatial search on road network. Zhang et al.~\cite{DBLP:conf/edbt/ZhangZZLCW14} studied the problem of diversified spatial keyword search on road networks. They designed an efficient signature-based inverted indexing to improve the search performance. Guo et al.~\cite{DBLP:journals/geoinformatica/GuoSAT15} studied continuous top-$k$ spatial keyword queries on road networks for the first time. They proposed two approaches which can monitor moving queries in an incremental manner and improve the search performance.

These aforementioned researches aim to find out spatial textual objects which are similar to the query. But they do not take into account the situation that query and objects containing geo-tagged images. Consequently, these approaches are not adequately suitable to overcome the challenge of region of visual interests query which contains geo-images.

\noindent\textbf{Content-Based Image Retrieval.} Content-based image retrieval (CBIR for short)~\cite{DBLP:journals/pami/JingB08,DBLP:journals/tomccap/LewSDJ06,DBLP:conf/mm/WuWS13} aims to search for images through analyzing their visual contents, and thus image representation~\cite{DBLP:conf/mm/WanWHWZZL14,DBLP:journals/pr/WuWGL18}. In these years, CBIR has attracted more and more attentions in the multimedia~\cite{TC2018,DBLP:journals/pr/WuWLG18} and computer vision community~\cite{DBLP:journals/tnn/WangZWLZ17,NNLS2018}. Many techniques have been proposed to support efficient multimedia query and image recognition. Scale Invariant Feature Transform (SIFT for short)~\cite{DBLP:conf/iccv/Lowe99,DBLP:journals/ijcv/Lowe04} is a classical method to extract visual features, which transforms an image into a large collection of local feature vectors. SIFT includes four main step: (1)scale-space extrema detection; (2)keypoint localization; (3)orientation assignment; (4)Kkeypoint descriptor. It is widely applied in lots of researches and applications. For example, Ke et al.~\cite{DBLP:conf/cvpr/KeS04} proposed a novel image descriptor named PCA-SIFT which combines SIFT techniques and principal components analysis (PCA for short) method. Mortensen et al.~\cite{DBLP:conf/cvpr/MortensenDS05} proposed a feature descriptor that augments SIFT with a global context vector. This approach adds curvilinear shape information from a much larger neighborhood to reduce mismatches. Liu et al.~\cite{DBLP:journals/inffus/LiuLW15} proposes a novel image fusion method for multi-focus images with dense SIFT. This dense SIFT descriptor can not only be employed as the activity level measurement, but also be used to match the mis-registered pixels between multiple source images to improve the quality of the fused image. Su et al.~\cite{Su2017MBR} designed a horizontal or vertical mirror reflection invariant binary descriptor named MBR-SIFT to solve the problem of image matching. Nam et al.~\cite{DBLP:journals/mta/NamKMHCL18} introduced a SIFT features based blind watermarking algorithm to address the issue of copyright protection for DIBR 3D images. Charfi et al.~\cite{DBLP:journals/mta/CharfiTAS17} developed a bimodal hand identification system based on SIFT descriptors which are extracted from hand shape and palmprint modalities.

Bag-of-visual-words~\cite{DBLP:conf/iccv/SivicZ03,DBLP:journals/tnn/WangZWLZ17,DBLP:journals/corr/abs-1804-11013}(BoVW for short) model is another popular technique for CBIR and image recognition, which was first used in textual classification. This model is a technique to transform images into sparse hierarchical vectors by using visual words, so that a large number of images can be manipulated. Santos et al.~\cite{DBLP:journals/mta/SantosMST17} presented the first ever method based on the signature-based bag of visual words (S-BoVW for short) paradigm that considers information of texture to generate textual signatures of image blocks for representing images. Karakasis et al.~\cite{DBLP:journals/prl/KarakasisAGC15} presents an image retrieval framework that uses affine image moment invariants as descriptors of local image areas by BoVW representation. Wang et al.~\cite{DBLP:conf/mmm/WangWLZ13} presented an improved practical spatial weighting for BoV (PSW-BoV) to alleviate this effect while keep the efficiency.

These researches aforementioned greatly improved the performance of image retrieval and recognition, but they do not consider the images with geo-tags which are generated by smart devices and location-based services. Consequently, these techniques can not be applied to address the region of visual interest query problem. 
\section{Preliminaries}
\label{preliminaries}
In this section, we propose the definition of region of visual interests (RoVI for short) at the first time, then present the notion of region of visual interests query (RoVIQ for short) and the similarity measurement. Besides, we review the techniques of image retrieval which is the base of our work. Table~\ref{tab:notation} summarizes the notations frequently used throughout this paper to facilitate the discussion.

\begin{table}
	\centering
    \small
	\begin{tabular}{|p{0.18\columnwidth}| p{0.73\columnwidth} |}
		\hline
		\textbf{Notation} & \textbf{Definition} \\ \hline\hline
		~$\mathcal{D}_I$                                 & A given database of geo-tagged images                \\ \hline
        ~$I_k$                                           & The $k$-th geo-tagged images                 \\ \hline
        ~$I_k.\lambda$                                   & The geo-location component of $I_k$                 \\ \hline
        ~$I_k.\psi$                                      & The visual component of $I_k$                 \\ \hline
        ~$\mathcal{U}$                                   & A region of visual interest user dataset                \\ \hline
        ~$u_i$                                           & A retion of visual interest user in $\mathcal{U}$        \\ \hline
	  	~$u_i.G$                                         & The geographical information component of $u_i$         \\ \hline
        ~$u_i.V$                                         & The visual information component of $u_i$         \\ \hline
        ~$W$                                             & A weight set of visual words          \\ \hline
        ~$w(v)$                                          & The weight of visual word $v$          \\ \hline
        ~$\mathcal{Q}$                                   & A region of visual interests query        \\ \hline
        ~$\mathcal{R}$                                   & The result set of a query                  \\ \hline
        ~$GeoSim(\mathcal{Q},u)$                         & The geographical similarity between $\mathcal{Q}$ and $u$     \\ \hline
		~$VisSim(\mathcal{Q},u)$                         & The visual similarity between $\mathcal{Q}$ and $u$     \\ \hline
        ~$\Gamma_G$                                      & The geographical similarity threshold             \\ \hline
        ~$\Gamma_V$                                      & The visual similarity threshold             \\ \hline
        ~$\Omega(u,u')$                                  & The region intersection of $u$ and $u'$        \\ \hline
        ~$\Theta(u,u')$                                  & The region union of $u$ and $u'$        \\ \hline
        ~$Area(x)$                                       & The function to compute the area of $x$             \\ \hline
        ~$N$                                             & A node of quadtree          \\ \hline
	\end{tabular}
    \caption{The summary of notations} \label{tab:notation}	
\end{table}

\subsection{Problem Definition}
\begin{definition}[\textbf{Geo-Image}] \label{def:geo-image}
Let $\mathcal{D}_I = \{I_1,I_2,...,I_{|\mathcal{D}_I|}\}$ be a database of geo-images in which $|\mathcal{D}_I|$ is the number of geo-images in it, each $I_i \in \mathcal{D}_I$ is an image with a geo-tag which is the geo-location information about the shoot place. We denote a geo-image as $I_i = (I_i.\lambda,I_i.\psi)$, wherein $I_i.\lambda$ is the geo-location component and $I_i.\psi$ is the visual component.
\end{definition}

\begin{definition}[\textbf{Region of Visual Interests (RoVI)}] \label{def:rovi}
Let $\mathcal{U} = \{u_1,u_2,...,u_{|\mathcal{U}|}\}$ be a region of visual interests user dataset in which $|\mathcal{U}|$ be the number of users in it. Each user $u_i \in \mathcal{U}$ is denoted as $u_i = (u_i.G,u_i.V)$, where $u_i.G$ represents the geographical information component and $u_i.V$ is the visual information component. The former is a geographical region, here we use minimum bounding rectangle (MBR for short) to represent it. Let $p_t$ and $p_b$ be the top-left point and bottom-right point of a MBR, the $u_i.G$ can be denoted as $u_i.G = (p_t,p_b)$. The visual information component is a vector of visual words $u_i.V = \{v_1,v_2,...,v_{|V|}\}$, which is generated from the geo-images located in this region. Let $W = \{w(v_1),w(v_2),...,w(v_{|V|})\}$ be the weight set of visual words. In the paper hereafter, whenever there is no ambiguity, "region of visual interest user" is abbreviated to "user".
\end{definition}

\begin{definition}[\textbf{Region of Visual Interests Query (RoVIQ)}] \label{def:aviquery}
Given a region of visual interest users dataset $\mathcal{U} = \{u_1,u_2,...,u_{|\mathcal{U}|}\}$, a region of visual interest query $\mathcal{Q} = (\mathcal{Q}.G,\mathcal{Q}.V)$ aims to return a set of users $\mathcal{R}$ from $\mathcal{U}$, which are similar to $\mathcal{Q}$, where $\mathcal{Q}.G$ denotes geographical region information and $\mathcal{Q}.V$ represents visual information. In formal, $\forall u \in \mathcal{R}$ and $\forall u' \in \mathcal{U} \setminus \mathcal{R}$,
\begin{equation*}
GeoSim(\mathcal{Q},u) \geq GeoSim(\mathcal{Q},u')
\end{equation*}
and,
\begin{equation*}
VisSim(\mathcal{Q},u) \geq VisSim(\mathcal{Q},u')
\end{equation*}
where $GeoSim(\mathcal{Q},u)$ and $VisSim(\mathcal{Q},u)$ are the geographical similarity function and visual similarity function respectively. To facilitate the computation, two thresholds $\Gamma_G$ and $\Gamma_V$ are defined, which represent geographical and visual similarity threshold respectively, and $\Gamma_G \in [0,1]$, $\Gamma_V \in [0,1]$. Therefore, for each user $u$ in results set $\mathcal{R}$,
\begin{equation*}
GeoSim(\mathcal{Q},u) \geq \Gamma_G
\end{equation*}
and,
\begin{equation*}
VisSim(\mathcal{Q},u) \geq \Gamma_V
\end{equation*}
consequently, the a RoVIQ $\mathcal{Q}$ is to return a set
\begin{equation*}
\mathcal{R} = \{u|GeoSim(\mathcal{Q},u) \geq \Gamma_G, VisSim(\mathcal{Q},u) \geq \Gamma_V, \forall u \in \mathcal{U}\}
\end{equation*}
\end{definition}

In order to measure the geographical similarity, we propose two important conceptions i.e., region intersection and region union in the following part.

\begin{definition}[\textbf{Region Intersection}] \label{def:interection}
Given a region of visual interest users dataset $\mathcal{U} = \{u_1,u_2,...,u_{|\mathcal{U}|}\}$, $\forall u, u' \in \mathcal{U}$, the region intersection of these two users is defined as $\Omega(u,u')$, denoted as follows:
\begin{equation*}
\Omega(u,u') = Area(u.G \bigcap u'.G)
\end{equation*}
where the function $Area(x)$ aims to compute the area of $x$. Similarly, for a RoVIQ $\mathcal{Q}$ and an user $u$, we can compute the region intersection of them as follows:
\begin{equation*}
\Omega(\mathcal{Q},u) = Area(\mathcal{Q}.G \bigcap u'.G)
\end{equation*}
\end{definition}

\begin{definition}[\textbf{Region Union}] \label{def:union}
Given a region of visual interest users dataset $\mathcal{U} = \{u_1,u_2,...,u_{|\mathcal{U}|}\}$, $\forall u, u' \in \mathcal{U}$, the region union of these two users is defined as $\Theta(u,u')$, denoted as follows:
\begin{equation*}
\Theta(u,u') = Area(u.G \bigcup u'.G)
\end{equation*}
Like the Definition~\ref{def:interection}, we can compute the region union of a query $\mathcal{Q}$ and an user $u$ as follows:
\begin{equation*}
\Theta(\mathcal{Q},u) = Area(\mathcal{Q}.G \bigcup u.G)
\end{equation*}
\end{definition}

\begin{definition}[\textbf{Geographical Similarity}] \label{def:geosim}
Given a region of visual interest users dataset $\mathcal{U} = \{u_1,u_2,...,u_{|\mathcal{U}|}\}$ and a RoVIQ $\mathcal{Q}$, $\forall u \in \mathcal{U}$, the geographical similarity between $\mathcal{Q}$ and $u$ is defined by
\begin{equation}\label{equ:geosim}
GeoSim(\mathcal{Q},u) = \frac{\Omega(\mathcal{Q},u)}{\Theta(\mathcal{Q},u)}
\end{equation}
\end{definition}

\begin{definition}[\textbf{Visual Similarity}] \label{def:vissim}
Given a region of visual interest users dataset $\mathcal{U} = \{u_1,u_2,...,u_{|\mathcal{U}|}\}$ and a RoVIQ $\mathcal{Q}$, $\forall u \in \mathcal{U}$, the visual similarity between $\mathcal{Q}$ and $u$ is measured by Jaccard similarity measurement, shown as below:
\begin{equation}\label{equ:vissim}
VisSim(\mathcal{Q},u) = \frac{\sum_{v \in \mathcal{Q}.V \bigcap u.V}^{}w(v)}{\sum_{v \in \mathcal{Q}.V \bigcup u.V}^{}w(v)}
\end{equation}
where $w(v)$ denotes the weight of visual word $v$.
\end{definition}

\begin{figure*}
\newskip\subfigtoppskip \subfigtopskip = -0.1cm
\begin{minipage}[b]{1\linewidth}
\begin{center}
     \includegraphics[width=1\linewidth]{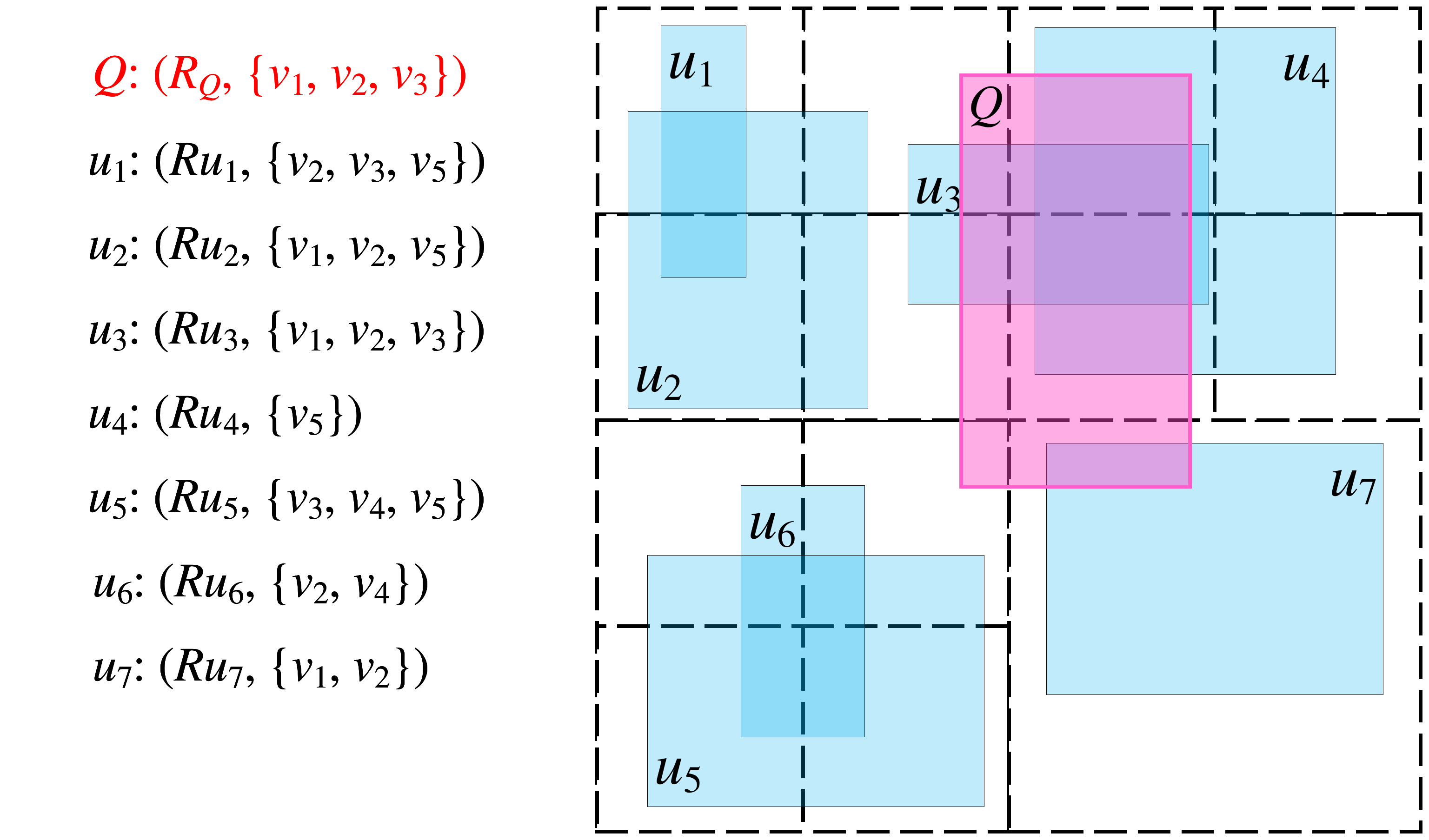}
   \captionsetup{justification=centering}
       \vspace{-0.2cm}
\caption{An example of region 0f visual interest query with users $\{u_1,u_2,...,u_7\}$ and a query $\mathcal{Q}$}
\label{fig:region-example}
\end{center}
\end{minipage}
\label{fig:k}
\end{figure*}

\begin{example}
\label{ex:example_query}
Figure~\ref{ex:example_query} shows an example of region of visual interest query. There are 7 users which contain geographical information denoted as rectangle $R_u$ and visual words set $\{v_1,v_2,...\}$. The light blue rectangles are the region of these users and the pink rectangle is the region of query.
\end{example}

\section{Baseline Method}
\label{base}

Before proceeding to present the proposed solution, we introduce how to exploit existing techniques to address the problem defined above. Three hybrid indexes, namely double index, visual first index, and spatial first index, are introduced in the following.

\subsection{Double Index}
\label{double}
\begin{figure*}
\newskip\subfigtoppskip \subfigtopskip = -0.1cm
\begin{minipage}[b]{0.99\linewidth}
\begin{center}
     \includegraphics[width=1\linewidth]{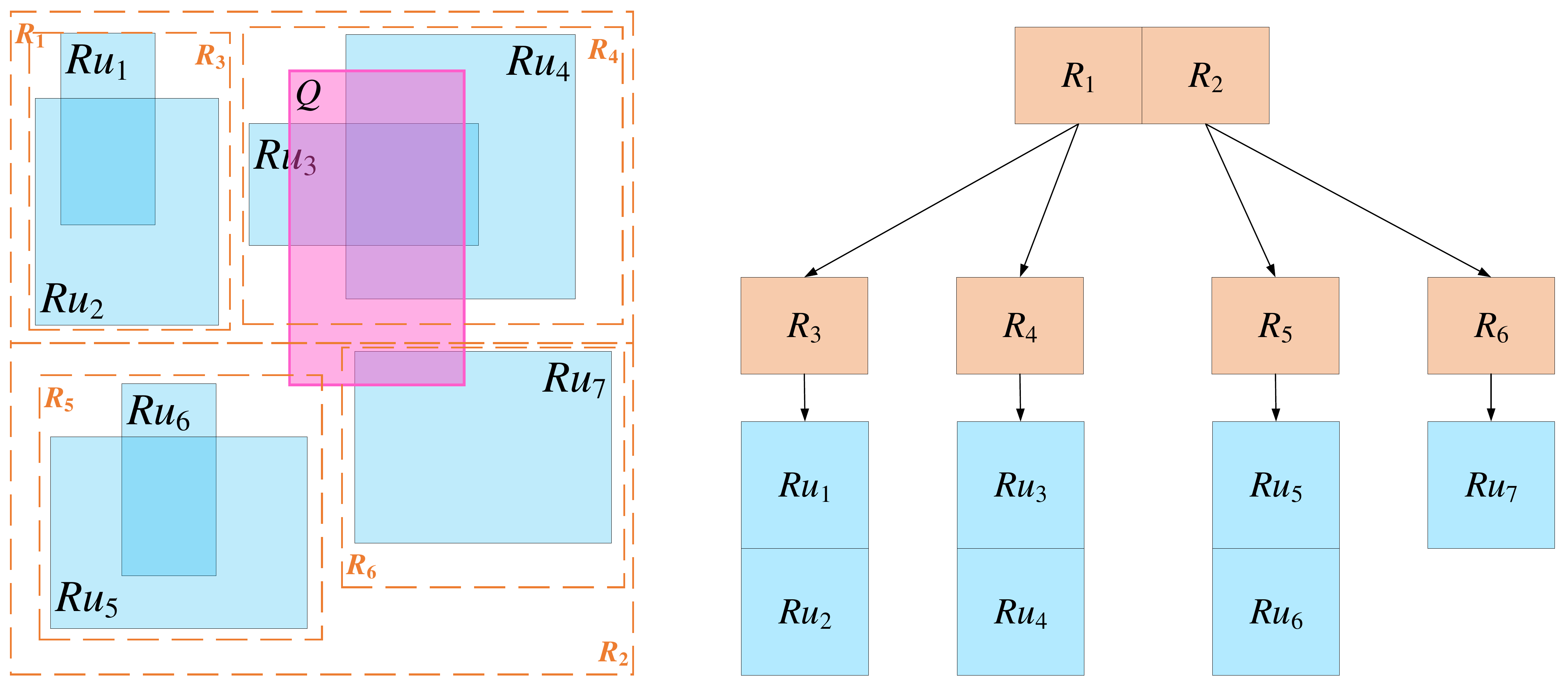}
   \captionsetup{justification=centering}
       \vspace{-0.2cm}
\caption{An example of double Index}
\label{fig:double-index}
\end{center}
\end{minipage}
\label{fig:fig}
\end{figure*}

Double index consists of two components: R*-tree and visual inverted files separately. For each user $u$, its geographical part is indexed by R*-tree and its visual information is indexed by inverted files. More specifically, the difference from conventional R*-tree is that the leaf node is composed of a series of region rather than a series of points. For example, in Figure.~\ref{fig:double-index} R*-tree contains two inner nodes $R_1$ and $R_2$. $R_1$ contains the elements $R_3$ and $R_4$, while $R_2$ contains the elements $R_5$ and $R_6$. And each leaf nodes include a set of Users. Visual inverted files are the same to conventional inverted files. It consists of a vocabulary for all distinct visual words in a collection of images and a set of posting lists related to this vocabulary. Each
posting list is a sequence of visual pairs $<id, w_{I,v}>$, where $id$ refers to
the user that contains visual words $v$, and $w_{I,v}$ is the weight of term
$v$ in image $I$.

When processing region of visual interests query, R*-tree is used to prune the irrelevant nodes as early as possible to shrink the search space, and visual inverted files are used
only the users containing at least one query visual word will be included in the search.
Because the candidates retrieved from  R*-tree only satisfy the spatial constraint, and the candidates obtained from visual inverted file only satisfy the visual word requirment. To answer region of visual interests query, the final result are the merge of the candidates from two indexes.

\begin{example}
We use an example to illustrate how double index works. Consider the users and region of visual interests query $\mathcal{Q}$ with thresholds $\Gamma_G$ = 0.3 and $\Gamma_T$ = 0.4 in Figure~\ref{fig:double-index}. Given a query $\mathcal{Q}$, the index first loads the root nodes of R*-tree, and gradually finds the intersection leaf node $R_4, R_6$. Then, we compute the geographical similarity between $U_3, U_4$ and $U_7$. As $GeoSim(\mathcal{Q},U_7) < \Gamma_G$, we discards $U_7$, and only add $U_3, U_4$ to candidate list. Then, we probes the visual inverted lists of $v_1, v_2$ and $v_3$ and finds the candidate users which satisfying $VisSim(\mathcal{Q},U) \geq \Gamma_V$, i.e., $U_1, U_2, U_3$ and $U_7$. Finally, we merge the candidate lists from R*-tree and visual inverted index, the list $U_3$ is reported as the final result.
\end{example}

\subsection{Visual First Index}
\label{vfi}
\begin{figure*}
\newskip\subfigtoppskip \subfigtopskip = -0.1cm
\begin{minipage}[b]{0.99\linewidth}
\begin{center}
     \includegraphics[width=1\linewidth]{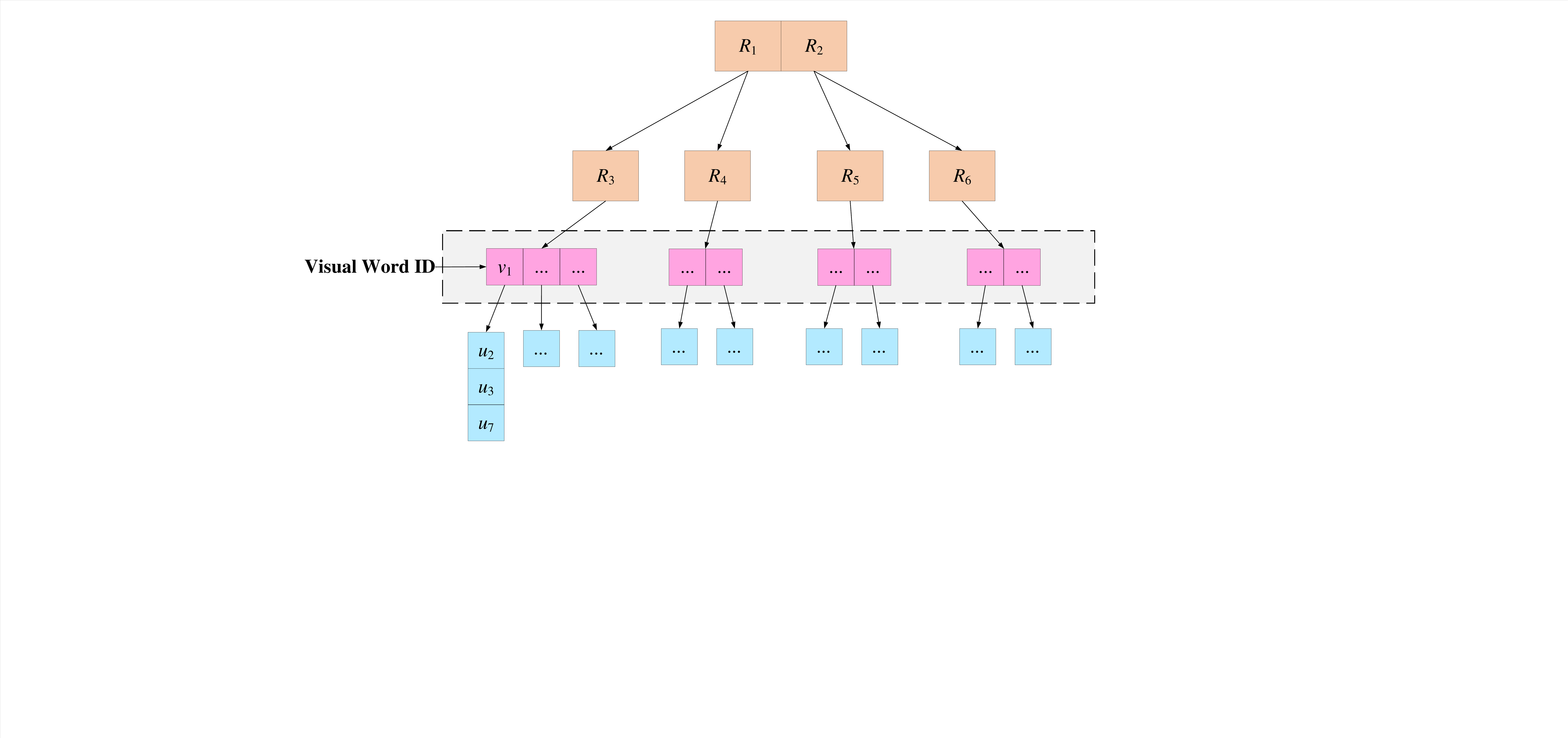}
   \captionsetup{justification=centering}
       \vspace{-0.2cm}
\caption{An example of visual first index}
\label{fig:vfi}
\end{center}
\end{minipage}
\label{fig:fig}
\end{figure*}

To efficiently facilitate the region of visual interests search, visual first index is fairly natural to employ the spatial index techniques to organize the users for each visual word, instead of keeping them in a list, as shown in Figure~\ref{fig:vfi}. For a given region of visual interests query, only the corresponding R*-trees related to query visual words need to access. Then, we can apply the region intersection operation on these R*-trees until arriving leaf level. For all users in leaf node, we first add the users with $VisSim(\mathcal{Q},U) \geq \Gamma_V$ to candidate list. Then, we further verify whether it satisfied spatial constraint, if $GeoSim(\mathcal{Q},U) \geq \Gamma_G$, we consider it as a result, otherwise we discard it.

\begin{example}
We use an example to illustrate how visual first index works. Consider the users and region of visual interests query $\mathcal{Q}$ with thresholds $\Gamma_G$ = 0.3 and $\Gamma_T$ = 0.4 in Figure~\ref{fig:vfi}. Given a query $\mathcal{Q}$, the index first loads the root nodes of $R*-tree_v1$, $R*-tree_v2$ and $R*-tree_v3$, find the users whose region intersect with query region in these three R*-tree and satisfying $VisSim(\mathcal{Q},U) \geq \Gamma_V$, i.e., $U_1, U_2, U_3$ and $U_7$. Then, for these users, we further verify whether it satisfied spatial constraint. Obviously, only $U_3$ satisfies the spatial requirement and can be considered as final result.
\end{example}

\subsection{Spatial First Index}
\label{sfi}
\begin{figure*}
\newskip\subfigtoppskip \subfigtopskip = -0.1cm
\begin{minipage}[b]{0.99\linewidth}
\begin{center}
     \includegraphics[width=1\linewidth]{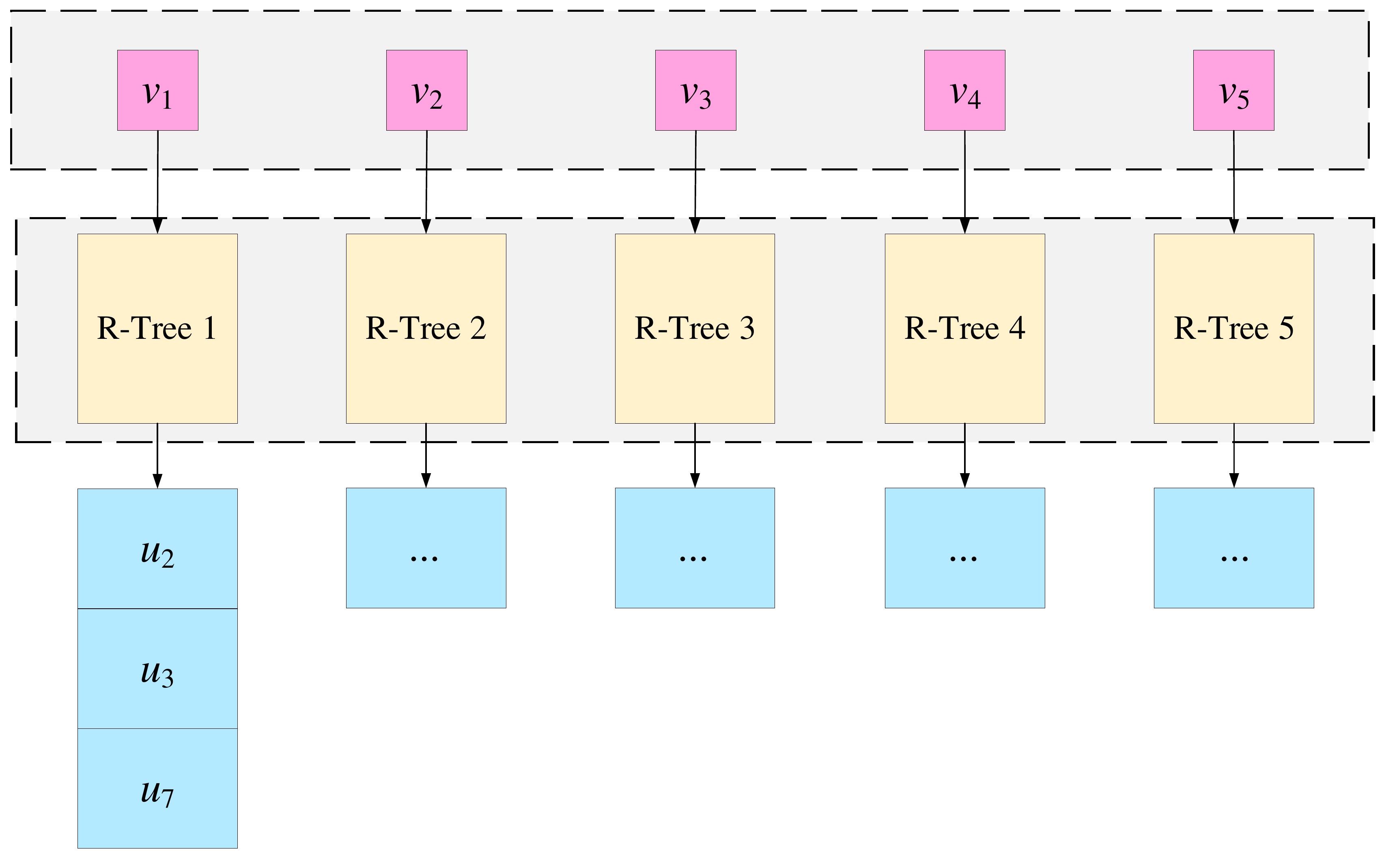}
   \captionsetup{justification=centering}
       \vspace{-0.2cm}
\caption{An example of spatial first index}
\label{fig:visual-keyword-index}
\end{center}
\end{minipage}
\label{fig:fig}
\end{figure*}

As shown in Figure~\ref{fig:visual-keyword-index}, a R*-tree is first built on all MBRs included in spatial scopes of all user' images. Next, all the users in each R*tree leaf node are indexed by visual inverted files based on their visual words. Hence, spatial first index consists of a primary R*-tree and  a series of secondary visual inverted files corresponding to R*-tree leaf nodes.

When processing a region of visual interests query, the geographical region of query is first used to find out the leaf nodes that may contain the candidates. If the leaf node intersect with the query region and $GeoSim(\mathcal{Q},N) \geq \Gamma_G$,
then the visual words of query are employed to load the corresponding visual inverted files based on this node, otherwise the leaf node are thrown away. For all users in leaf node, we judge each user whether $GeoSim(\mathcal{Q},U) \geq \Gamma_G$, if it satisfied the spatial constraint, we add the users to candidate list. Then, we further verify whether it satisfied visual requirement, if $VisSim(\mathcal{Q},U) \geq \Gamma_V$, we add it to result list, otherwise we discard it.

\begin{example}
We use an example to illustrate how spatial first index works. Consider the users and region of visual interests query $\mathcal{Q}$ with thresholds $\Gamma_G$ = 0.3 and $\Gamma_T$ = 0.4 in Figure~\ref{fig:visual-keyword-index}. Given a query $\mathcal{Q}$, the R*-tree index first finds all leaf nodes intersect with the query region. i.e., $R_4$ and $R_6$. As $GeoSim(\mathcal{Q},R_6) < \Gamma_G$, only leaf node $R_4$ needs to further treatment. Then, we load the visual inverted index of $R_4$ for $v_1$, $v_2$, and $v_3$. Both $U_3$ and $U_4$ satisfied spatial constraint, but only $U_3$ satisfied visual requirement. Thus, the final result is $U_3$.
\end{example}

\section{Quadtree Based Inverted Visual Index}
\label{spatialindex}

In this section, we present a novel indexing technique named quadtree based inverted visual index based on shared quadtree and inverted index for the region of visual interest search problem. In subsection~\ref{gqtree}, we present our spatial indexing structure, and in subsection~\ref{rovisearch}, we propose the efficient algorithm named RoVI Search to address the problem of RoVI query based on this novel index.

\subsection{Spatial Indexing Structure}
\label{gqtree}

\subsubsection{Virtual Quadtree}

\begin{figure*}
\newskip\subfigtoppskip \subfigtopskip = -0.1cm
\begin{minipage}[b]{1\linewidth}
\begin{center}
     \includegraphics[width=0.8\linewidth]{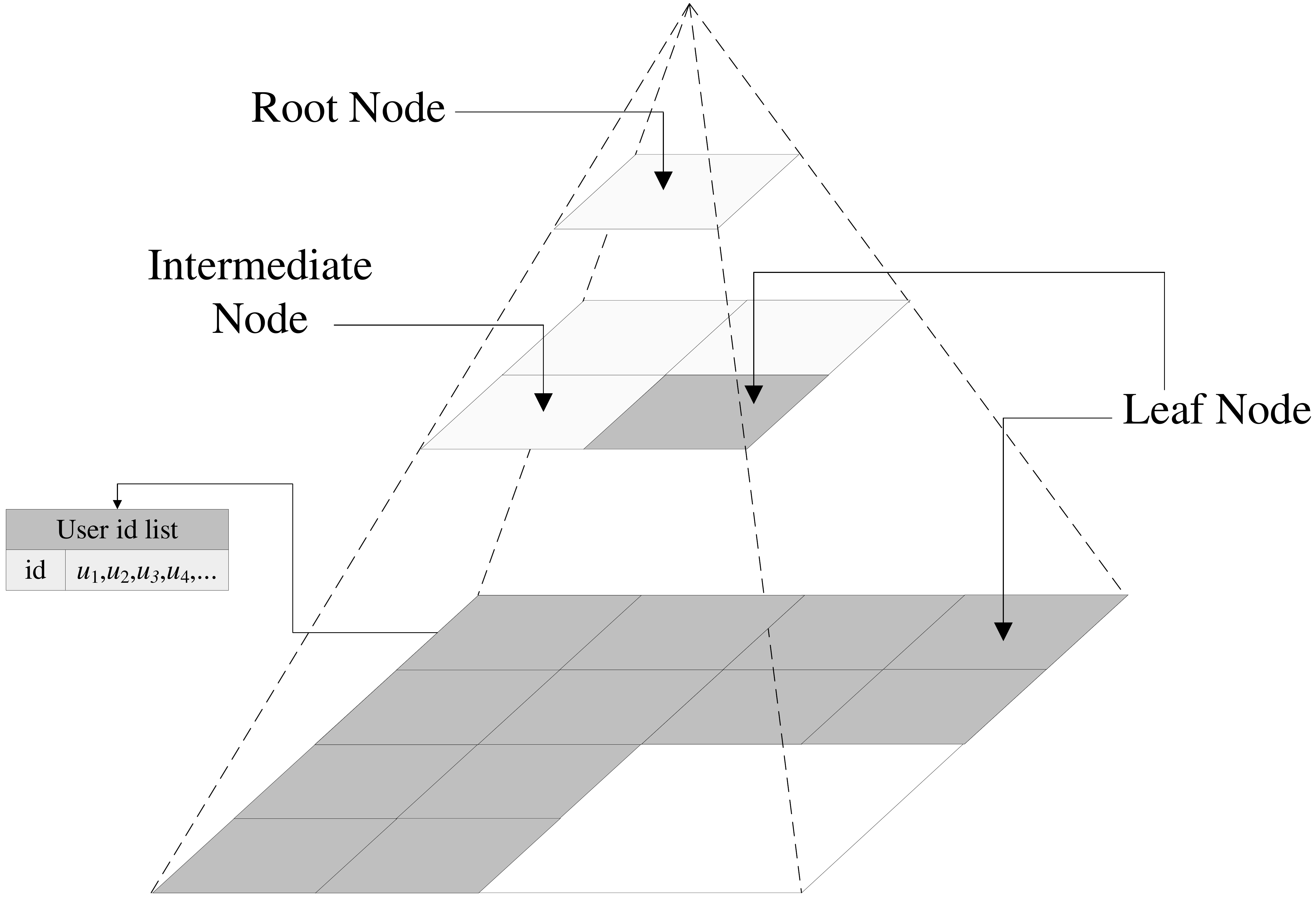}
   \captionsetup{justification=centering}
       \vspace{-0.2cm}
\caption{An example of a virtual quadtree}
\label{fig:fig3}
\end{center}
\end{minipage}
\label{fig:k}
\end{figure*}

We utilize shared quadtree to construct our index. Quadtree~\cite{DBLP:journals/cacm/Gargantini82,DBLP:journals/pami/HunterS79} is a classical space-partitioning indexing technique which is to subdivide a $d$-dimensional space into 2$d$ regions in a recursive manner. Figure~\ref{fig:fig3} illustrate a virtual quadtree. It decomposes the space into $L$ levels and for $l$-th level, the space is split into 4$^l$ equal area region. Each node in quadtree represents a geographical region. The root node of the virtual quadtree represents the entire geographical region, which is corresponding to level 0. For level 1, there are four equal area nodes which are partitioned from the root node, and the rest can be done in the same manner. There are three colors of nodes shown in Figure~\ref{fig:fig3}. Specifically, the root node and intermediate node are colored by light gray, which are already split into four subnodes. The dark gray nodes denote leaf nodes which can be located in any level of the tree according to the split condition. The node with white color in level 2 is not maintained in fact. Each node has a list of users whose the geo-location are in this node, but these users containing both geographical information and visual information are stored in the leaf node, which is maintained by an user id list. To sum up, the whole geographical region is spatially indexed into several nodes, and the users distribute in them.

Inserting a user $u$ into a virtual quadtree can be operated in a traditional manner which travels the whole quadtree from root node to find the leaf node whose geographical region denoted as $N.G$ overlaps the geographical area of the user, i.e., $u.G$. If the geographical region of a leaf node overlaps the region of user $u$, i.e., $GeoSim(u.G,N.G) \geq 0$, then $u$ can be inserted into $N$.

\subsubsection{Z-order curve}

Each node of the virtual quadtree can be encoded by the space filling curve techniques, such as Morton code~\cite{Morton2015A}, Hilbert code and gray code~\cite{Faloutsos1986Multiattribute}. For this study, we encode the quadtree nodes according to Morton code, i.e., Z-order curve, since it is encoded based on its partition sequence. The Z-order curve describe the path of the node in the virtual quadtree.

\begin{figure*}
\newskip\subfigtoppskip \subfigtopskip = -0.1cm
\begin{minipage}[b]{1\linewidth}
\begin{center}
     \subfigure[Z-order curve]{
     \includegraphics[width=0.48\linewidth]{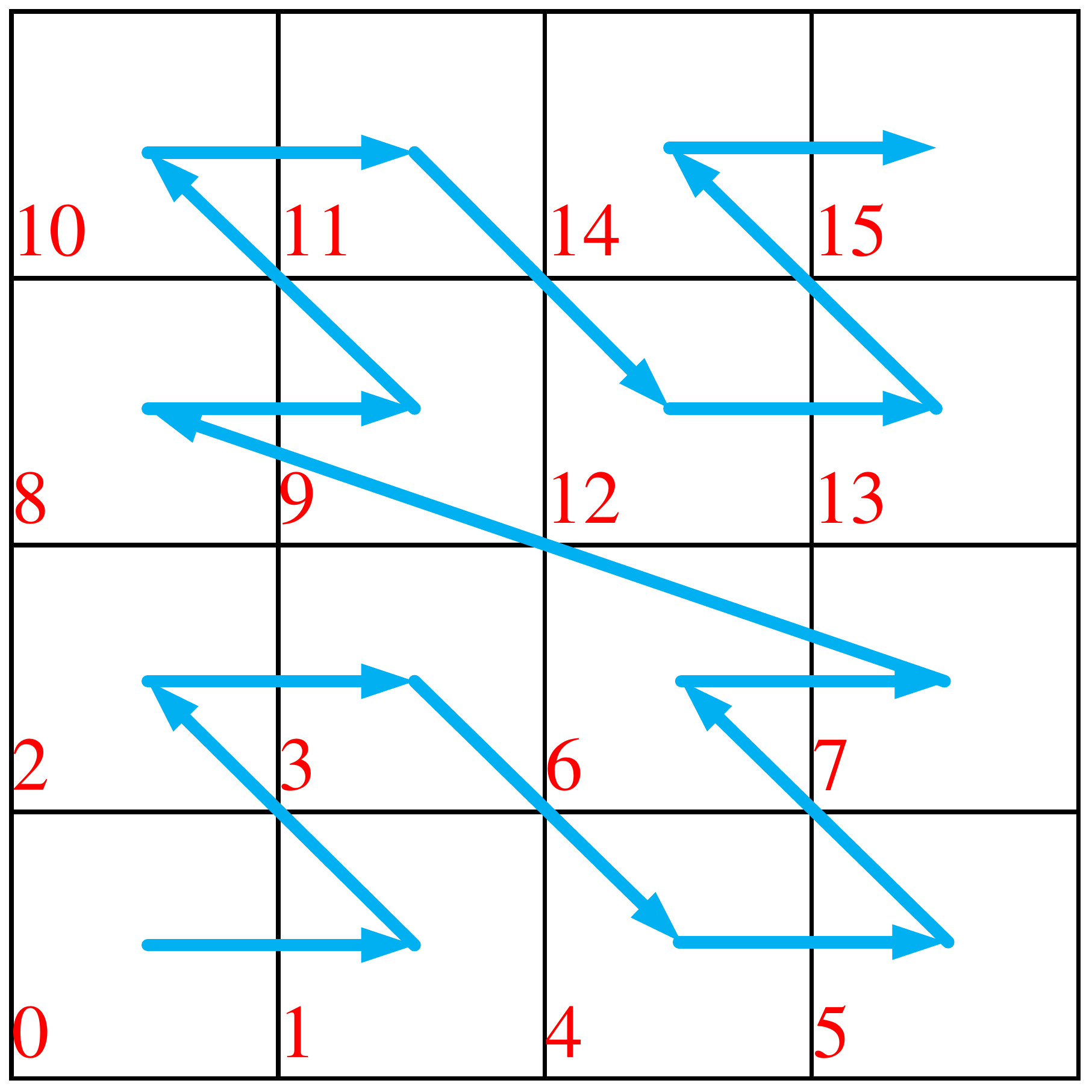}
     }
     \subfigure[encoding of nodes by z-order]{
     \includegraphics[width=1\linewidth]{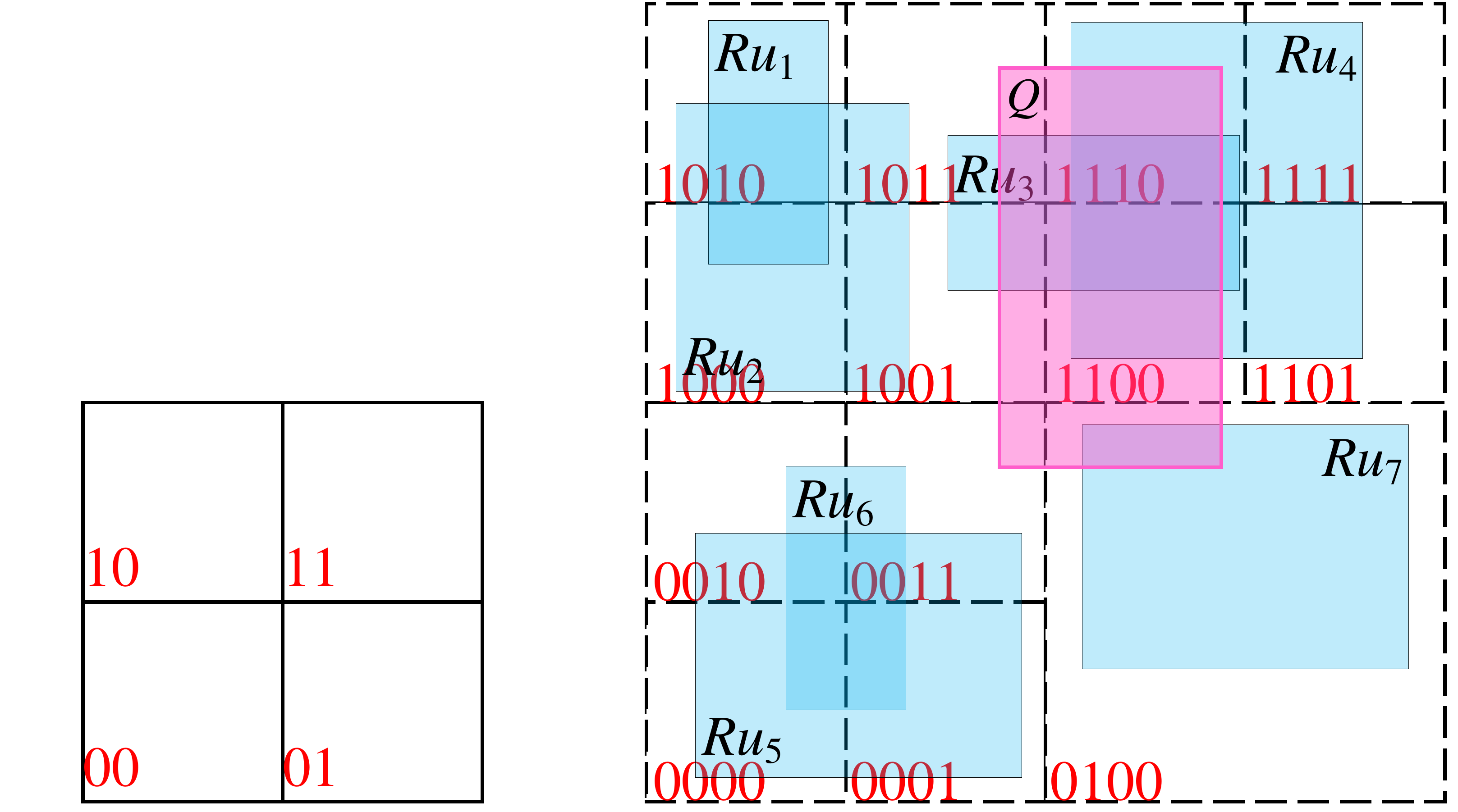}
     }
   \captionsetup{justification=centering}
       \vspace{-0.2cm}
\caption{Z-order and partition}
\label{fig:fig4}
\end{center}
\end{minipage}
\end{figure*}

Figure.~\ref{fig:fig4}(a) demonstrate the how to obtain the Morton code of a node according to the partition sequences in 2-dimensional space. We assume that a node is split into four subnodes in the order as of Z-order curve. We denote these four subnodes by Z-order curve as 1, 2, 3, 4 respectively. For the situation shown in Figure.~\ref{fig:fig4}(a), the sixteen node are coded from 0 to 15 in decimal.

Figure.~\ref{fig:fig4}(a) shows that the our encoding method of quadtree node by Z-order. For level 1 of quadtree, there are four subnode, we encode them in the form of binary, i.e., 00, 01, 10, 11 as the order of z-order curve. For sixteen subnotes in level 2, the code of them are from 0000 to 1111. In general, for the $l$-th level, the code of each node has 2$^l$ bits. Thus, in the example shown by Figure.~\ref{fig:fig4}(b), the code of node in level 2 has 4 bits. On the other hand, this example illustrates an example of distribution of 7 users $\{u_1,u_2,u_3,u_4,u_5,u_6,u_7\}$ in the whole geographical region colored by light blue and a query $\mathcal{Q}$ colored by pink. According to the code of each node, we can generate a region list for each user. For instance, for user $u_1$, the region list is denoted as $\mathcal{L}_{u_1} = \{N_8,N_{10}\}$. The region list of $u_2$ is $\mathcal{L}_{u_2} = \{N_8,N_9,N_{10},N_{11}\}$.

\subsubsection{Quadtree Based Inverted Visual Index}

\begin{figure*}
\newskip\subfigtoppskip \subfigtopskip = -0.1cm
\begin{minipage}[b]{1\linewidth}
\begin{center}
     \includegraphics[width=1\linewidth]{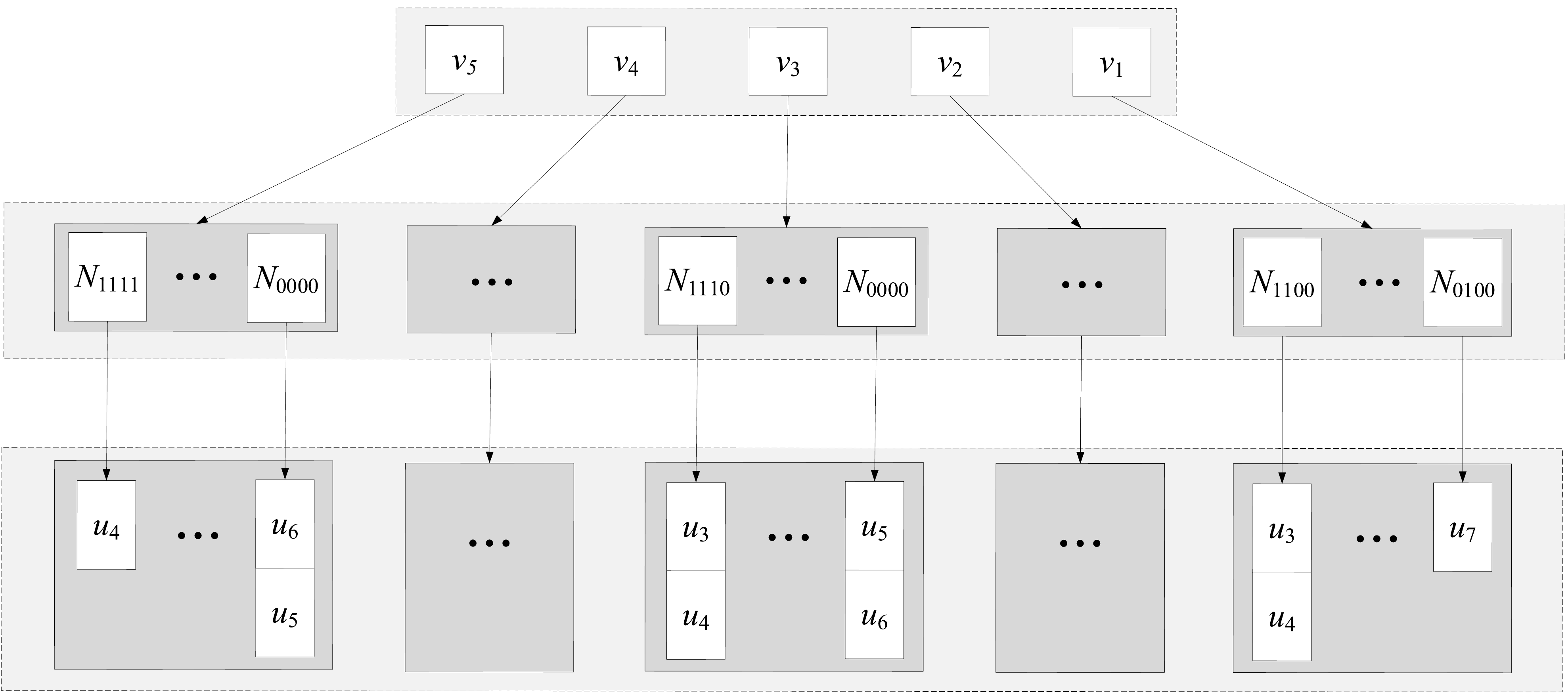}
   \captionsetup{justification=centering}
       \vspace{-0.2cm}
\caption{An example of a quadtree based inverted visual index}
\label{fig:fig5}
\end{center}
\end{minipage}
\label{fig:k}
\end{figure*}

We propose a novel inverted index structure based on quadtree mentioned above and visual words. This inverted index structure has three layers: the first layer is a visual words list which contains the visual words generated from all of the geo-images in users. Each of it contains a node list pointer which is point to a node list in the second layer. The node list is constructed by the visual quadtree introduced above based on z-order code, in which each node has an user list pointer. Third layer consists of several user lists. The user list pointer of a node is to point to the user list in which all the users are included in this node. All the user lists are stored in disk. Figure~\ref{fig:fig5} illustrates a novel inverted index of example~\ref{ex:example_query}. There are 5 visual word generated from 7 users, which form the first layer. For visual word $v_1$, there are three users containing it, i.e., $u_2$, $u_3$ and $u_7$, and $u_7$ is in node $N_{0100}$, $u_3$ and $u_4$ are in node $N_{1100}$.

\noindent\textbf{Visual Filter}. We design a visual filter for candidate set generation during the RoVI Search. According to the visual similarity measurement defined in Section~\ref{preliminaries}, we can define a candidate visual threshold $\mathfrak{c}_V = \Gamma_V * \sum_{v \in \mathcal{Q}.V}^{}w(v)$. It is clearly that visual similarity between a query and a user $VisSim(\mathcal{Q}.V,u.V) \geq \Gamma_V$ only if $\sum_{v \in \mathcal{Q}.V \bigcap u.V}^{}w(v) \geq \mathfrak{c}_V$. Therefore, we can use the threshold $\mathfrak{c}_V$ as a visual filter.

\subsection{RoVI Search Algorithm Based on Quadtree Based Inverted Visual Index}
\label{rovisearch}

According to quadtree based inverted visual index, we design an efficient search algorithm for RoVI Query. The pseudo-code is show as follows.

\begin{algorithm}
\begin{algorithmic}[1]
\footnotesize
\caption{\bf RoVI Search Algorithm}
\label{alg:rovisearch}

\INPUT  A RoVI query $\mathcal{Q}$, a quadtree $\mathcal{T}$, visual similarity threshold $\Gamma_V$, geographical similarity threshold $\Gamma_G$.
\OUTPUT A results set $\mathcal{R}$.

\STATE Initializing: $\mathcal{R} \leftarrow \emptyset$; //A results set
\STATE Initializing: $\mathcal{N} \leftarrow \emptyset$; //The set of nodes
\STATE Initializing: $L_N \leftarrow \emptyset$; //A node list
\STATE Initializing: $\hat{\mathcal{N}} \leftarrow \emptyset$; // The set of intersect nodes
\STATE Initializing: $\mathcal{C} \leftarrow \emptyset$; The candidates set

\STATE $\hat{\mathcal{N}} \leftarrow GetInterectNodes(\mathcal{T}.Root,\mathcal{Q})$;
\FOR{each $v$ in $\mathcal{Q}.V$}
    \STATE $N \leftarrow GetNode(\mathcal{T}.Root,v)$;
    \STATE $\mathcal{N}$.Add($N$);
\ENDFOR
\FOR{each node $\hat{N}$ in $\hat{\mathcal{N}}$}
    \STATE $N \leftarrow OverlapNode(\hat{N},\mathcal{N})$;
    \IF{$VisualFilter(\mathcal{Q},N) \geq \mathfrak{c}_V$}
        \STATE $L_N.Add(N)$;
    \ENDIF
\ENDFOR
\STATE $\mathcal{C} \leftarrow LoadCandidateUsers(L_N)$;
\FOR{each user $u$ in $\mathcal{C}$}
    \STATE $t_V \leftarrow VisSim(\mathcal{Q}.V,u.V)$;
    \STATE $t_G \leftarrow GeoSim(\mathcal{Q}.G,u.G)$;
    \IF{$t_V \geq \Gamma_V$ and $t_G \geq \Gamma_G$}
        \STATE $\mathcal{R}.Add(u)$;
    \ENDIF
\ENDFOR
\RETURN $\mathcal{R}$;
\end{algorithmic}
\end{algorithm}

Algorithm~\ref{alg:rovisearch} illustrates the RoVI Search processing. Firstly the virtual quadtree is loaded and the algorithm find out the intersect nodes by the query $\mathcal{Q}$ and save them into a set $\hat{\mathcal{N}}$. Then for each visual word contained in query, search out the node which contains any of these visual words and stores them into $\mathcal{N}$. For each node in the set of intersect nodes, the algorithm find out all the nodes which is overlap with it in $\mathcal{N}$, and then the $VisualFilter()$ can be called to filter these nodes which are stored in node list $L_N$. The next step is to generate the candidate set by calling the procedure $LoadCandidateUsers(L_N)$. For each candidate, the algorithm computes the visual similarity and geographical similarity by execute $VisSim(\mathcal{Q}.V,u.V)$ and $GeoSim(Q.G,u.G)$. If these two similarities are all greater than or equal to the threshold, then $u$ will be added into the results set.  
\section{PERFORMANCE EVALUATION}
\label{perform}

In this section, we present results of a comprehensive performance study on real
image datasets to evaluate the efficiency and scalability of the proposed techniques.
Specifically, we evaluate the effectiveness of the following indexing techniques for region of visual interests search on road network.
\begin{itemize}
\item{\textbf{DI}} DI is the technique proposed in ~\ref{double}.
\item{\textbf{VFI}} VFI is the technique proposed in ~\ref{vfi}.
\item{\textbf{SFI}} SFI is the technique proposed in ~\ref{sfi}.
\item{\textbf{QIV}} QIV is quadtree based inverted visual index technique, which are proposed in Section  ~\ref{spatialindex}.
\end{itemize}

\noindent \textbf{Datasets.} Performance of various algorithms is evaluated on both real spatial and image datasets.
The following two datasets are deployed in the experiments. Real dataset \textbf{Flickr} is obtained by crawling millions image the photo-sharing site Flickr(\url{http://www.flickr.com/}). To evaluate the scalability of our proposed algorithm, The dataset size varies from 20K to 1M. The spatial locations of \textbf{Flickr} is
obtained from the US Board on Geographic Names(\url{http://geonames.usgs.gov}). We selected 1 million
POIs from this dataset as centers and extended the POIs
with random widths and heights to generate regions. Similarly, Real dataset \textbf{ImageNet} is obtained from is the largest image dataset ImageNet, which is widely used in image processing and computer vision. it includes 14,197,122 images and 1.2 million images with SIFT features. We generate \textbf{ImageNet} dataset with varying size from 20K to 1M. The region of the users are randomly generated from spatial datasets Rtree-Portal (
\url{http://www.rtreeportal.org}) with the same method as dataset \textbf{Flickr}.

{\bf Workload.}
A workload for the region of visual interests query consists of $100$ queries. The  query response time is employed to evaluate the performance of the algorithms. The image dataset size grows from 0.2M to 1M; the number of the query visual words changes from 50 to 150;
the spatial similarity and visual similarity varies from 0.1 to 0.5. To generate a query region, we randomly select from the locations of the above spatial dataset, then the query region is set to a rectangle centered at these selected users' locations. The query region varies from 1\% of the data space to 5\% of the data space. By default,  The image dataset size, the number of the query visual words, the spatial similarity, the visual similarity, and the query region set to 1M, 100, 0.3, 0.3 and 2\% respectively. Experiments are run on a PC with Intel Xeon 2.60GHz dual CPU and 16G memory running Ubuntu. All algorithms in the experiments are implemented in Java. Note that the virtual quadtree of QIV is maintained in memory, because the index only occupied very small memory space.

\begin{figure*}
\newskip\subfigtoppskip \subfigtopskip = -0.1cm
\begin{minipage}[b]{1\linewidth}
\begin{center}
     \subfigure[Evaluation on Flickr]{
     \includegraphics[width=0.48\linewidth]{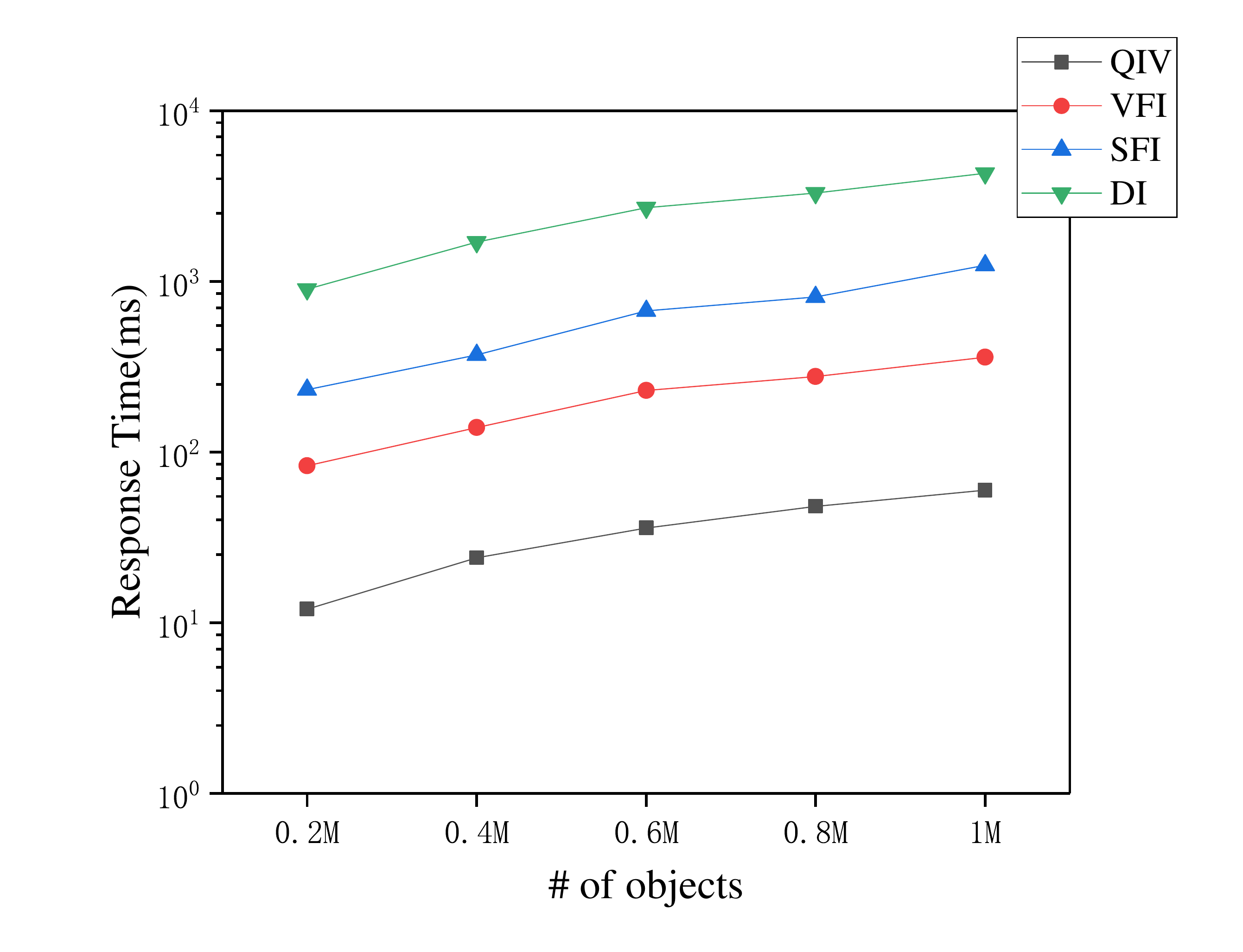}
     }
     \subfigure[Evaluation on ImageNet]{
     \includegraphics[width=0.48\linewidth]{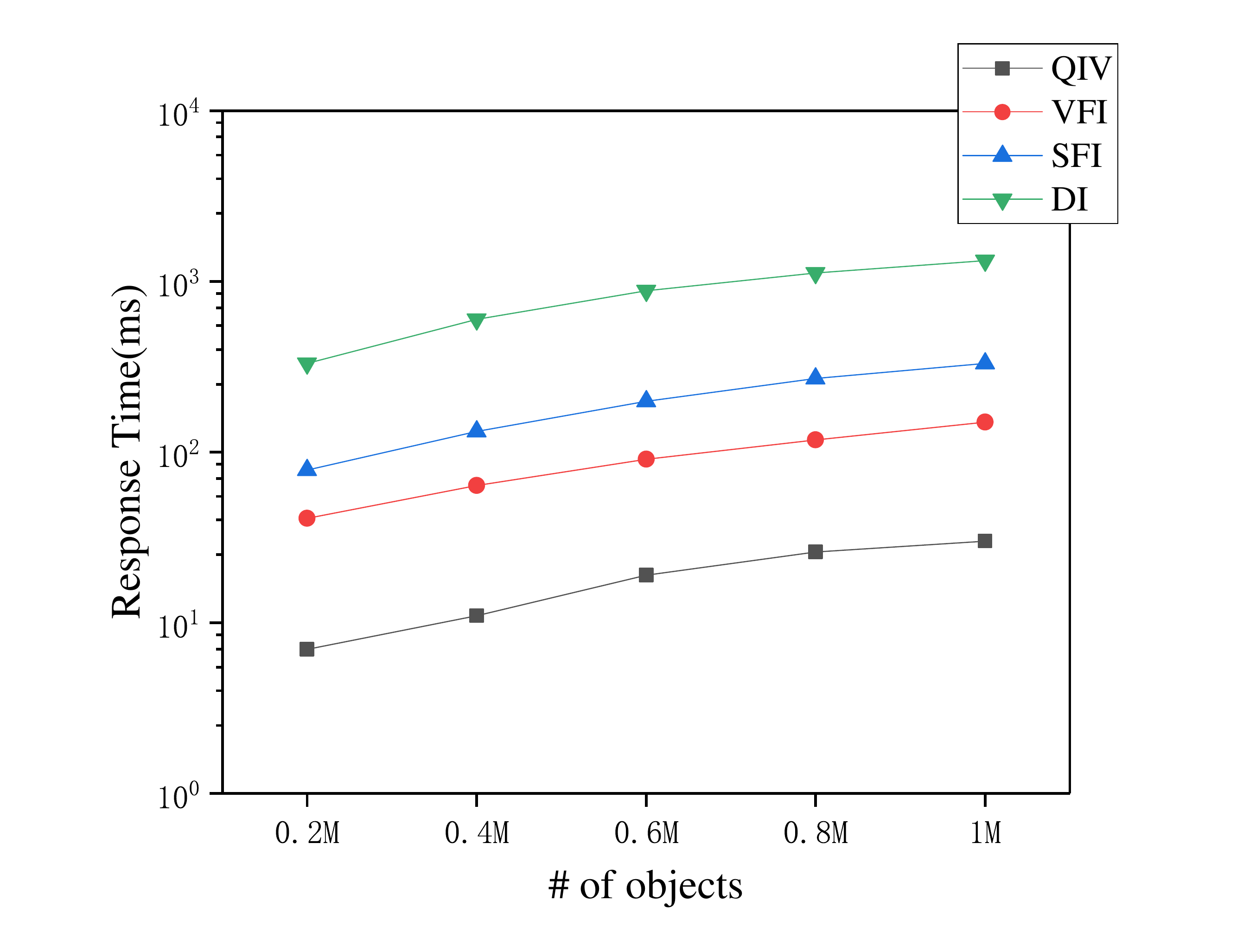}
     }
   \captionsetup{justification=centering}
       \vspace{-0.2cm}
\caption{Evaluation on the number of objects on Flickr and ImageNet}
\label{fig:number-of-objects}
\end{center}
\end{minipage}
\end{figure*}

\noindent\textbf{Evaluation on the size of dataset.} We evaluate the effect of the size of dataset on Flickr and ImageNet shown in Figure~\ref{fig:number-of-objects}. It is clear that the response time of QIV, VFI, SFI, DI rise with the increase of size of dataset on both two datasets. Figure~\ref{fig:number-of-objects}(a) illustrates that the performance of our method, QIV, is much higher than others. The time cost of DI is the largest of them, it gradually increase from about 1000 ms to 4500 ms around. The response time of VFI and SFI are less than DI, both of them shows a fluctuating upward from 0.2M to 1M. Figure~\ref{fig:number-of-objects}(b) shows the experiment on ImageNet. Like the evaluation on Flickr, the time cost of QIV is the lowest of these four methods. It shows a moderate growth with the increment of size of dataset. Similarly, the response time of DI is the highest, it climbs slowly and finally hit 1000ms at 1M. Apparently, the gentle upward trend of VIF and SIF are very similar, they are obvious inferior to QIV.

\begin{figure*}
\newskip\subfigtoppskip \subfigtopskip = -0.1cm
\begin{minipage}[b]{1\linewidth}
\begin{center}
     \subfigure[Evaluation on Flickr]{
     \includegraphics[width=0.48\linewidth]{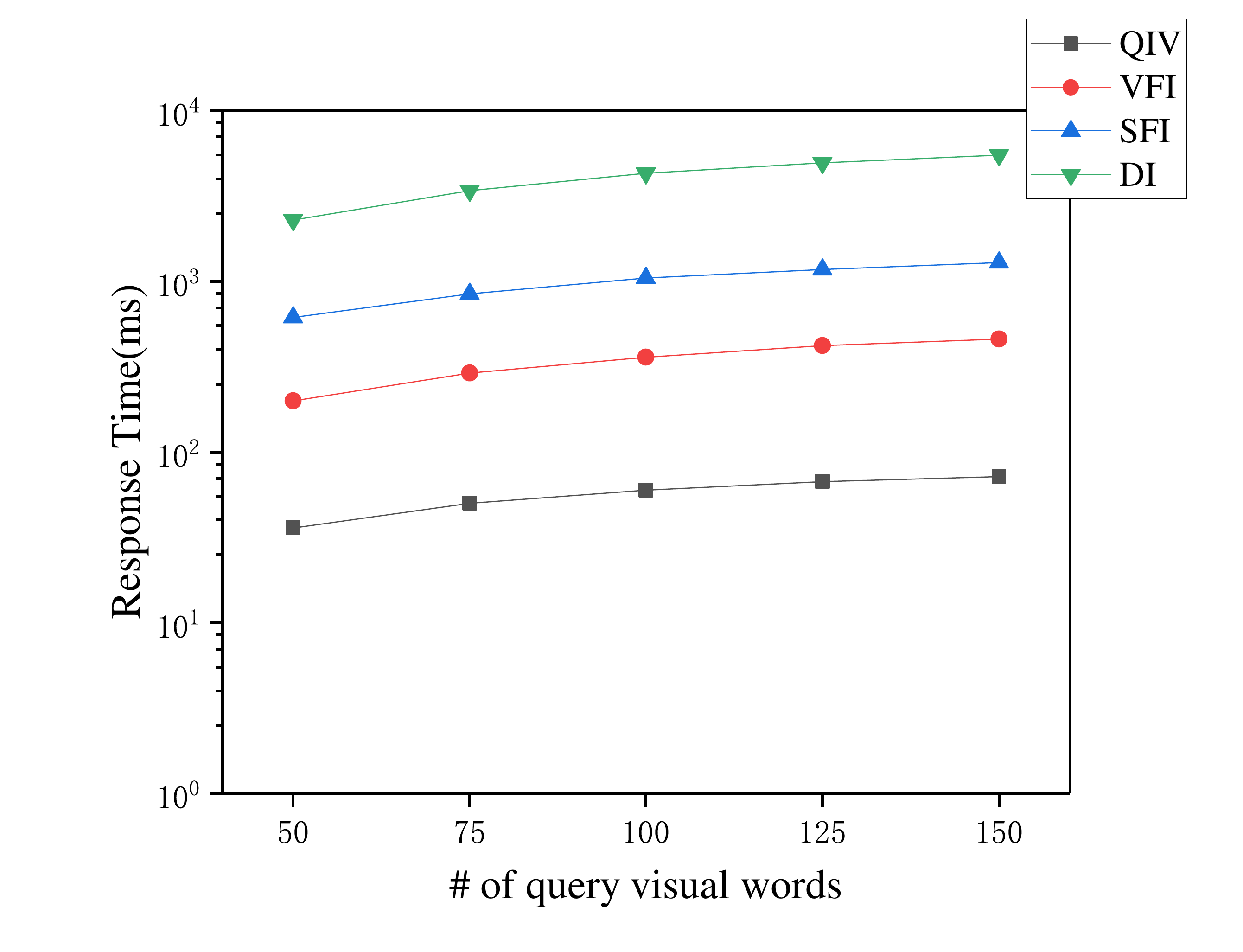}
     }
     \subfigure[Evaluation on ImageNet]{
     \includegraphics[width=0.48\linewidth]{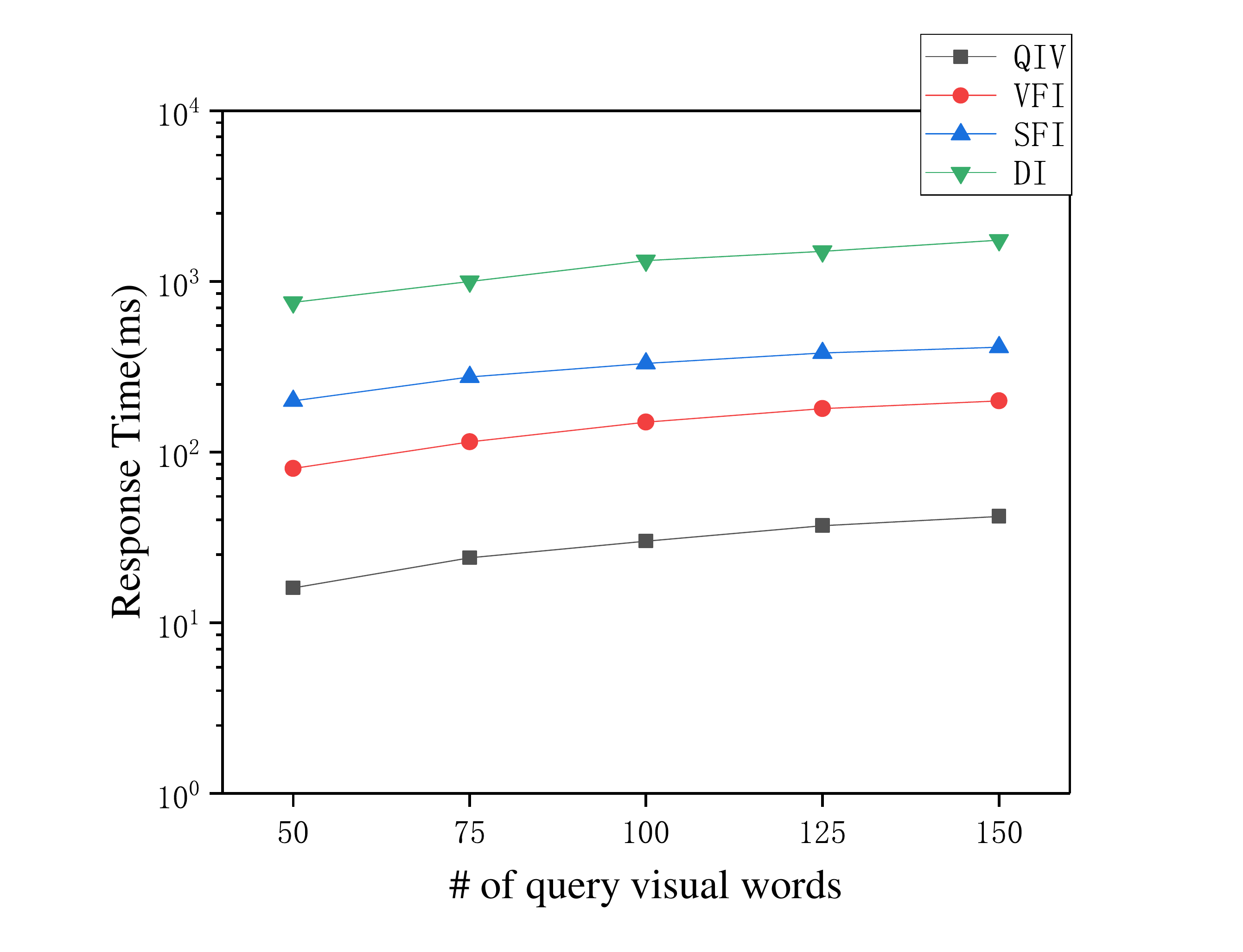}
     }
   \captionsetup{justification=centering}
       \vspace{-0.2cm}
\caption{Evaluation on the number of query visual words on Flickr and ImageNet}
\label{fig:number-of-query-visual-words}
\end{center}
\end{minipage}
\end{figure*}

\noindent\textbf{Evaluation on the number of query visual words.} We evaluate the effect of the number of query visual words on Flickr and ImageNet dataset shown in Figure~\ref{fig:number-of-objects}. The experiment on Flickr is shown in Figure~\ref{fig:number-of-objects}(a). We can see that the response time of them ascend with the rising of the number of query visual words. Specifically, the performance of our method is the best. It increase very slowly from 50 to 150. On the other hand, the time cost of DI is the highest. the growth trend of VFI and SFI are similar, both of them are increase smoothly. In Figure~\ref{fig:number-of-objects}(b), all of the trends are similar to the situation in Figure~\ref{fig:number-of-objects}(a). In other words, all of them climb gradually and the efficiency of QIV outperforms VFI, SFI and DI.

\begin{figure*}
\newskip\subfigtoppskip \subfigtopskip = -0.1cm
\begin{minipage}[b]{1\linewidth}
\begin{center}
     \subfigure[Evaluation on Flickr]{
     \includegraphics[width=0.48\linewidth]{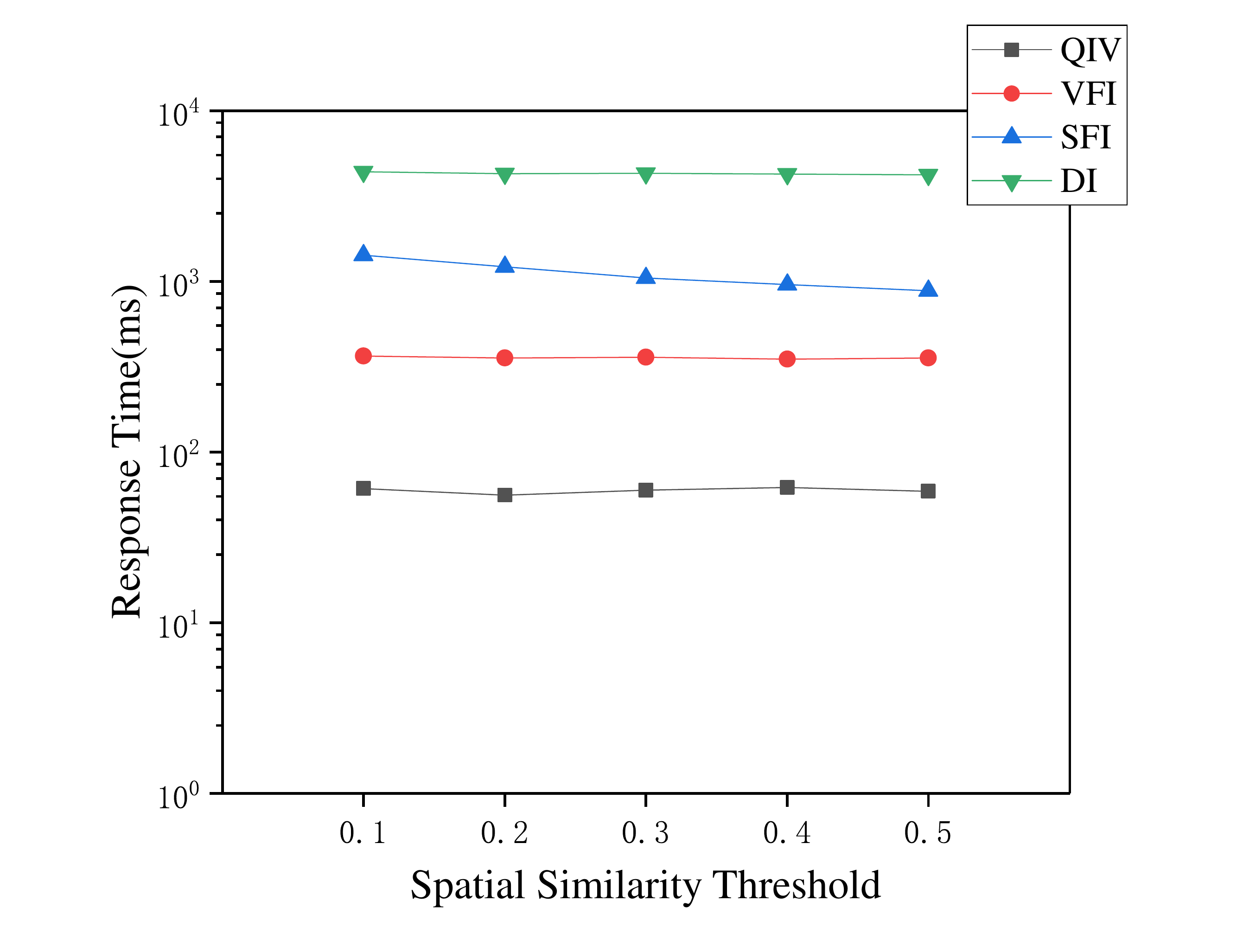}
     }
     \subfigure[Evaluation on ImageNet]{
     \includegraphics[width=0.48\linewidth]{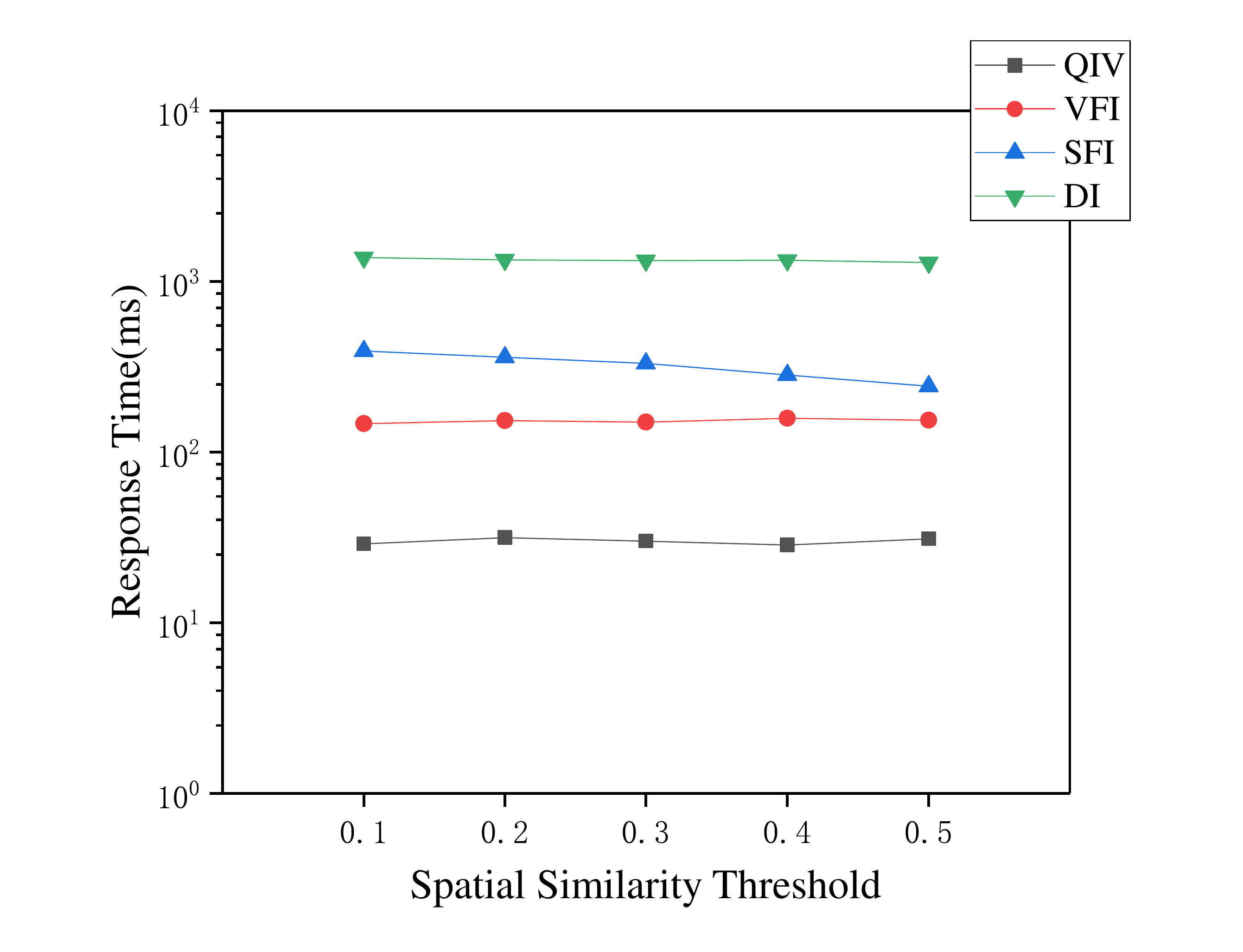}
     }
   \captionsetup{justification=centering}
       \vspace{-0.2cm}
\caption{Evaluation on the spatial similarity threshold on Flickr and ImageNet}
\label{fig:time-spatial-similarity-threshold}
\end{center}
\end{minipage}
\end{figure*}

\noindent\textbf{Evaluation on the spatial similarity threshold.} We evaluate the effect of the spatial similarity threshold on Flickr and ImageNet dataset shown in Figure~\ref{fig:time-spatial-similarity-threshold}. In Figure~\ref{fig:time-spatial-similarity-threshold}(a), with the increment of spatial similarity threshold, the response time of QIV moderately fluctuates around 80ms, which is the lowest in these methods. The trends of VFI and DI are almost unchanged, but the time cost of the latter is much higher than the former. On the other hand, the performance of SIF shows a moderate decrement with the rising of spatial similarity threshold, which is better than DI. Figure~\ref{fig:time-spatial-similarity-threshold}(b) shows that the efficiency of DI is almost the same with the increment of spatial similarity threshold, which is much lower than SFI, VFI and QIV. The response time of SFI decrease slowly, which is higher than VFI. Like the experiment on Flickr, our method has the best performance, which fluctuates slightly with the increasing of the threshold.

\begin{figure*}
\newskip\subfigtoppskip \subfigtopskip = -0.1cm
\begin{minipage}[b]{1\linewidth}
\begin{center}
     \subfigure[Evaluation on Flickr]{
     \includegraphics[width=0.48\linewidth]{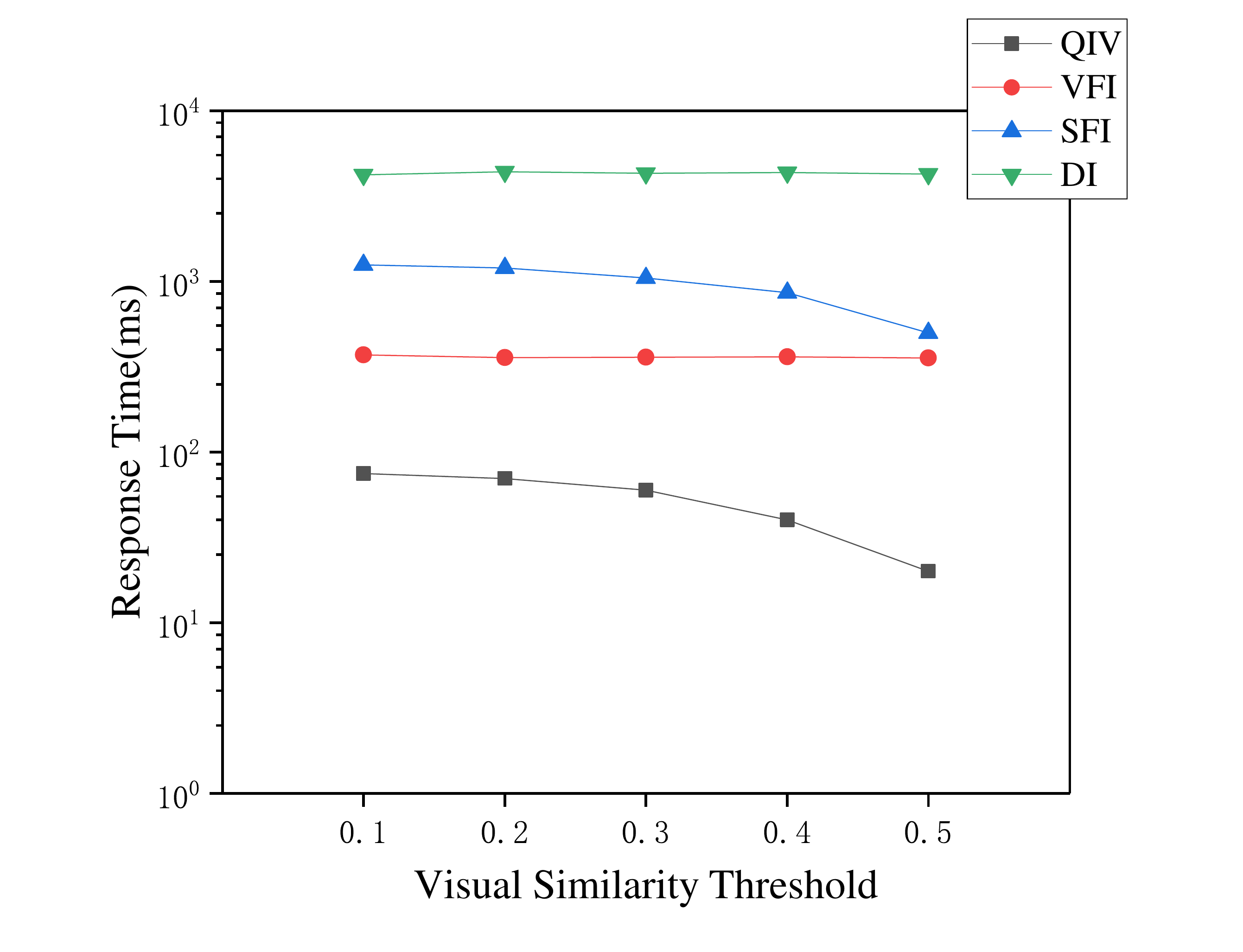}
     }
     \subfigure[Evaluation on ImageNet]{
     \includegraphics[width=0.48\linewidth]{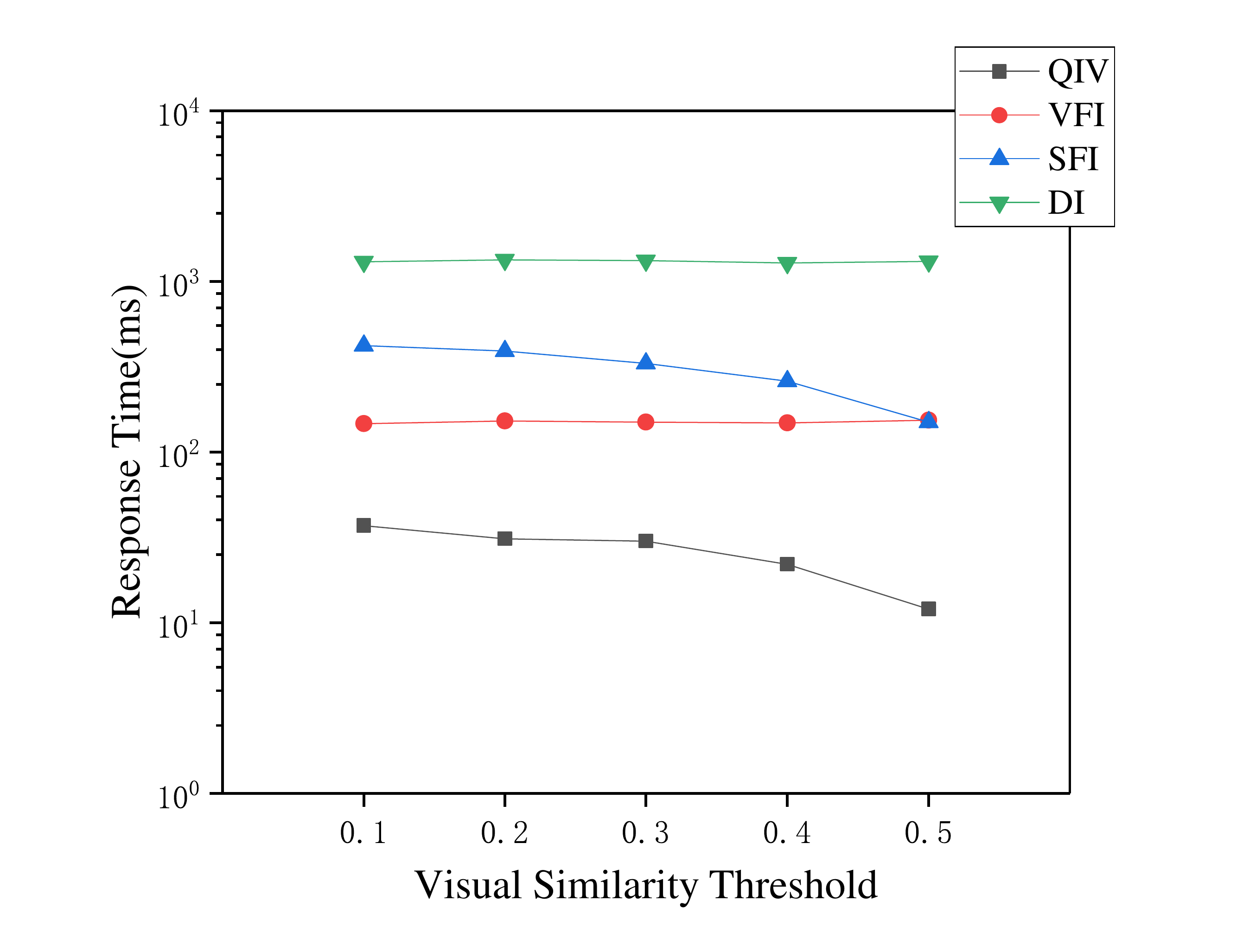}
     }
   \captionsetup{justification=centering}
       \vspace{-0.2cm}
\caption{Evaluation on the visual similarity threshold on Flickr and ImageNet}
\label{fig:time-visual-similarity-threshold}
\end{center}
\end{minipage}
\end{figure*}

\noindent\textbf{Evaluation on the visual similarity threshold.} We evaluate the effect of the visual similarity threshold on Flickr and ImageNet dataset shown in Figure~\ref{fig:time-spatial-similarity-threshold}. We can find from the Figure~\ref{fig:time-spatial-similarity-threshold}(a) that the performance of DI and VFI are almost unchanged with the increasing of visual similarity threshold. The former is the highest of them. By comparison, the time cost of VFI is much less than DI. On the other hand, our method QIV and SFI shows a decrement. In the interval of $[0.1,0.3]$, both of them go down very slowly and after that, they decrease obviously. Clearly, the performance of QIV is the best of them. Figure~\ref{fig:time-spatial-similarity-threshold}(b) illustrates that all of these treads are like the situation in Figure~\ref{fig:time-spatial-similarity-threshold}(a). QIV has the best performance, which decreases step by step with the rising of the threshold. The response time of SFI gradually goes down and at 0.5 it is a litter lower than VIF. Clearly, DI shows the worst efficiency on this dataset.

\begin{figure*}
\newskip\subfigtoppskip \subfigtopskip = -0.1cm
\begin{minipage}[b]{1\linewidth}
\begin{center}
     \subfigure[Evaluation on Flickr]{
     \includegraphics[width=0.48\linewidth]{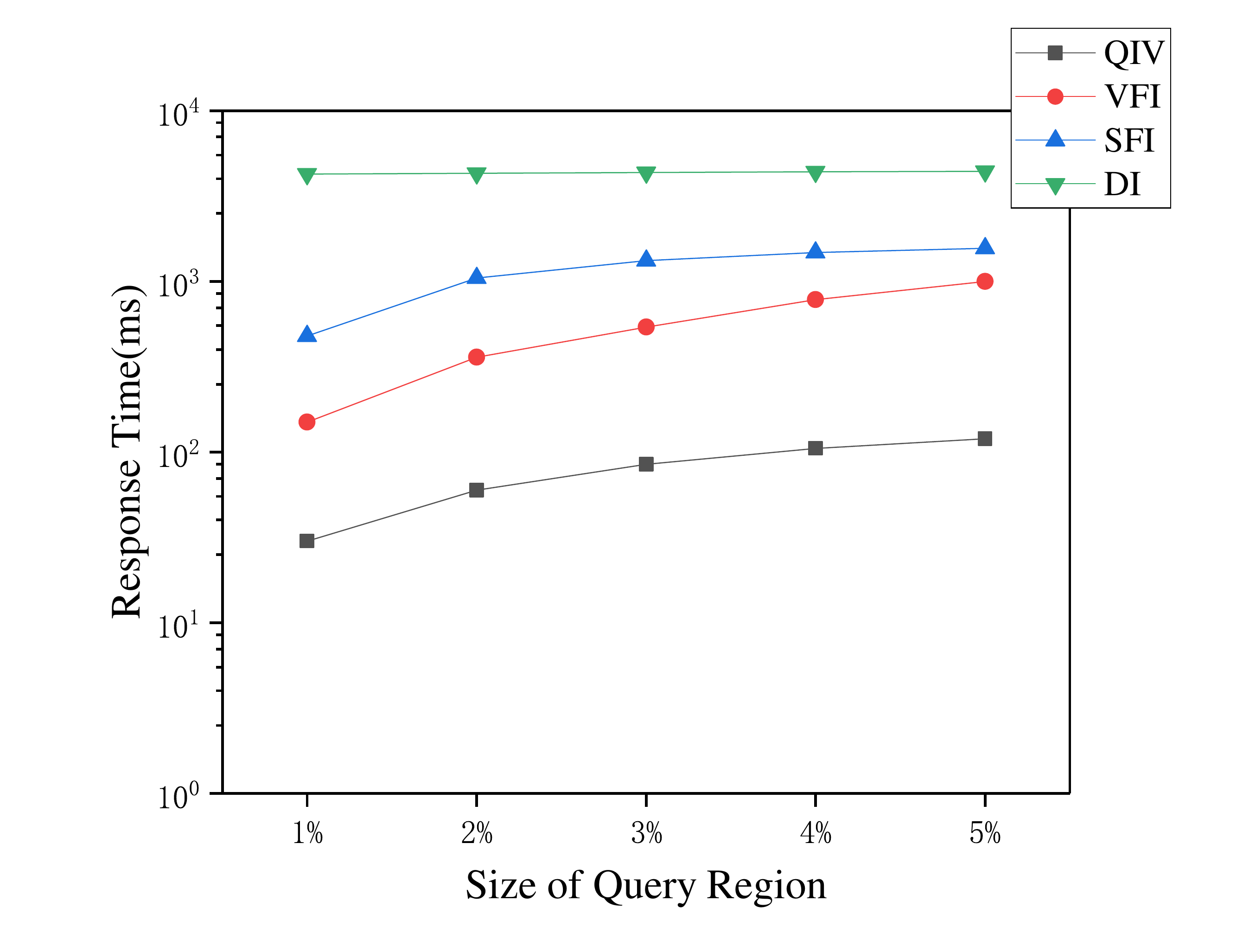}
     }
     \subfigure[Evaluation on ImageNet]{
     \includegraphics[width=0.48\linewidth]{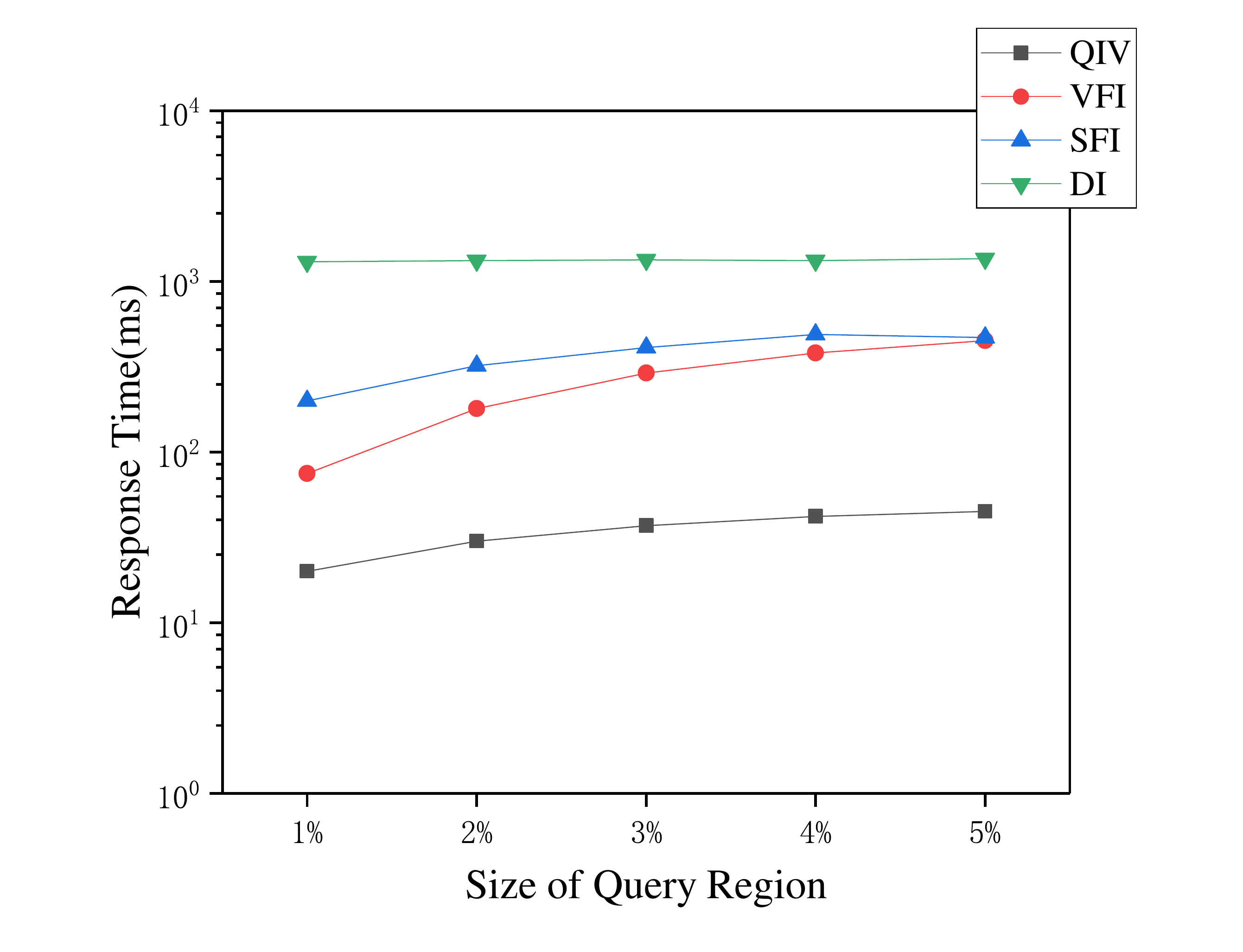}
     }
   \captionsetup{justification=centering}
       \vspace{-0.2cm}
\caption{Evaluation on the different size of query region on Flickr and ImageNet}
\label{fig:size-of-query-region}
\end{center}
\end{minipage}
\end{figure*}

\noindent\textbf{Evaluation on the visual similarity threshold.} We evaluate the effect of different size of query region on Flickr and ImageNet dataset shown in Figure~\ref{fig:size-of-query-region}. Figure!\ref{fig:size-of-query-region}(a) tells us that the response time of SFI, VFI and QIV increase step by step with the increasing of the query region size. In the interval $[1\%,3\%]$, the time cost of all these three methods rise evidently and after that the upward trends have become very gentle. The performance of DI is almost unaffected by the change of query region size, which is much lower than the performance of our method. In Figure~\ref{fig:size-of-query-region}(b), we find that the growth trends of SFI and VFI is obvious, and at 5\% they are almost the same. By comparison, the upward trend of QIV is moderate, and the performance of DI is almost the same with the increment of region size. On both these two dataset, our method shows the best performance.

\section{Conclusion}
\label{con}

In this paper, we study a novel query problem named region of visual interests (RoVI) query. Given a set of region of visual interests users which contains geographical information and visual information, a RoVI query aims to find out the users which are similar to the query in both aspects of geographical similarity and visual similarity. Firstly we define RoVI query in formal and then propose the geographical and visual similarity function. In order to improve the efficiency of searching, we design a novel spatial indexing technique called quadtree based inverted visual index and a efficient algorithm called region of visual interests search. Besides, we introduce three baselines to explain how to exploit existing techniques to address this problem. The experimental evaluation on real geo-multimedia dataset shows that our solution outperforms the state-of-the-art method.

\textbf{Acknowledgments:} This work was supported in part by the National Natural Science Foundation of China
(61702560), project (2018JJ3691, 2016JC2011) of Science and Technology Plan of Hunan Province, and the Research and Innovation Project of Central South University Graduate Students(2018zzts177,2018zzts588).

\bibliographystyle{spmpsci}      
\bibliography{ref}

\end{document}